 \documentclass[12pt]{iopart}
\usepackage[dvips]{graphicx}
\usepackage{cite} 
\usepackage{color}
\usepackage{calc}
\usepackage[utf8]{inputenc}

\usepackage{amssymb}

\begin{document}
\title{Performance of a $^{229}$Thorium solid-state nuclear clock}
    
\author{G.~A.~Kazakov$^{1,2*}$, A.~N.~Litvinov$^{2}$,V.~I.~Romanenko$^{3}$, 
L.~P.~Yatsenko$^{3}$, A.~V.~Romanenko$^{4}$, M.~Schreitl$^1$, G.~Winkler$^1$,
and T.~Schumm$^1$}

\address{$^{1}$ Institute for Atomic and Subatomic Physics, TU Wien, Stadionallee 2, 1020 Vienna, Austria}

\address{$^{2}$ St. Petersburg State Polytechnical University, 29, Polytechnicheskaya st, 
St. Petersburg 195251, Russia}

\address{$^{3}$ Institute of Physics, Nat. Acad. Sci. of Ukraine, 46, Nauky Ave., 
Kyiv 03028, Ukraine}

\address{$^{4}$ Kyiv National Taras Shevchenko University, 4, Academician Glushkov Ave, 
Kyiv 03022, Ukraine}

\ead{{kazakov.george@gmail.com}}
\begin{abstract}
The $7.8$\,eV nuclear isomer transition in $^{229}$Thorium has been suggested as a clock transition in a new type of optical frequency standard. Here we discuss the construction of a ``solid-state nuclear clock" from Thorium nuclei implanted into single crystals transparent in the vacuum ultraviolet range. We investigate crystal-induced line shifts and broadening effects for the specific system of Calcium fluoride. At liquid Nitrogen temperatures, the clock performance will be limited by decoherence due to magnetic coupling of the Thorium nuclei to neighboring nuclear moments, ruling out the commonly used Rabi or Ramsey interrogation schemes. We propose clock stabilization based on a flourescence spectroscopy method and present optimized operation parameters. Taking advantage of the high number of quantum oscillators under continuous interrogation, a fractional instability level of $10^{-19}$ might be reached within the solid-state approach.
\end{abstract}


\maketitle                


\section{Introduction}
Atomic clocks have a significant impact both on fundamental science as well as on everyday life technology. 
Prominent examples for the latter are satellite-based navigation systems, such as GPS, GLONASS, and GALILEO, and telecommunication networks. 
On the fundamental side, the current definition of the unit of time (SI second) is a duration corresponding to 9,192,631,770 periods of microwave (MW) radiation stablized to the ground state hyperfine transition in $^{133}$Cesium~\cite{Markowitz58}. The physical implementation is a global network of primary Cesium frequency standards, Cs atomic fountain clocks have reached an uncertainty level of $4.5 \cdot 10^{-16}$~\cite{Parker10}. In fountain clocks the interrogation time is limited by the ballistic flight time of the atoms within the apparatus, further progress within this approach can be achieved in space-based realizations as the PHARAO project~\cite{Lemonde99}.

The recent progress in atom cooling and trapping together with the emergence of sub-hertz linewidth lasers allows to develop more accurate frequency standards based on \emph{optical} transitions using \emph{trapped} neutral atoms and ions. 
For example, an inaccuracy level evaluation for a neutral $^{87}$Sr optical lattice clock is $1.5 \cdot 10^{-16}$~\cite{Ludlow08}, and the $^{27}$Al$^+$ quantum logic frequency standard demonstrated an inaccuracy level of the order of $10^{-17}$~\cite{Chou10}. These approaches benefit from a $10^4-10^5$ gain in quality factor by going from microwave to optical transitions and longer observation times using trapped quantum oscillators.

A promising candidate for the role of a clock transition in a next-generation quantum frequency standard is the nuclear isomer transition in $^{229}$Thorium. This unique radio isotope has an extremely low-energy $^{229m}$Th isomeric state. The most recent estimate of the transition energy is $7.8 \pm 0.6$\,eV, obtained from indirect measurements of the $^{233}$U$\rightarrow ^{229}$Th decay spectrum with an advanced X-ray microcalorimeter~\cite{Beck09}. The corresponding wavelength of about 160\,nm in the vacuum ultraviolet (VUV) is attainable by modern laser systems and hence the construction of a Thorium ``nuclear clock" seems plausible. 

An optical frequency standard based on a nuclear transition will open new possibilities for experimental studies of basic laws of physics, i.e. constraining possible drifts of fundamental constants of interactions. In~\cite{Litvinova09} it is conjectured that a possible drift of the fine structure constant $\delta \alpha/\alpha$ will result a corresponding but amplified change of the $^{229}$Th nuclear transition frequency $\delta \omega/\omega$ with an amplification factor between $10^2$ and $10^4$. Hence the Thorium quantum frequency standard will be several order of magnitude more sensitive to possible drifts of $\alpha$ than existing standards based on electronic shell transitions in atoms or ions.

It should be noted that the current energy value of the Thorium isomer transition is derived indirectly by looking at $\gamma$ transitions with energies much higher than the isomer one. Therefore this value can not be considered definitive until a direct measurement is performed. Several groups worldwide (i.e. PTB, Germany; Georgia Institute of Technology, USA;  University of California, Los Angeles, USA; Vienna University of Technology, Austria) are currently working towards this goal. 

Two main experimental approaches towards a Thorium clock, first proposed in \cite{Peik03}, are under realization now: the use of trapped Thorium ions~\cite{Peik09, PorsevPeik10, Campbell09} and a ``solid-state nuclear spectroscopy''~\cite{Schumm, Rellergert10}.

In the first scheme, individual or small ensembles (up to $10^5$) of ions are trapped and laser cooled using established techniques. The main advantage of the trapped ion approach is a precise knowledge of and control over the fields affecting the ions. It has been proposed that a nuclear clock based on a single trapped $^{229}$Th$^{3+}$ ion might reach a fractional inaccuracy of $10^{-19}$~\cite{Campbell11}.

In the second approach, a macroscopic number (up to $10^{18}$) of Thorium atoms is implanted into a VUV-transparent crystal and interrogated simultaneously. This scheme will require none of the complex experimental techniques for ion trapping and cooling and can be fabricated in mass production. The number of simultaneously excited Thorium nuclei is many orders of magnitudes higher than in the trapped ion approach, therefore the solid-state spectroscopy method seems to be more promising for an initial localization of frequency of the isomer transition.

In the present work we analyze the possibility of constructing a ``solid-state nuclear clock'' based on $^{229}$Thorium-doped crystals.  Here, certain difficulties arise from the systematic crystal effects and the very long relaxation time of the excited nuclear state that can significantly exceed the interrogation time. We briefly review the current state of knowledge on the Thorium isomer energy and lifetime and consider transition line shifts, splittings, and broadening processes due to the crystal environment for the specific case of Thorium doped into Calcium flouride (CaF$_2$) crystals (section~\ref{sec:nucleus}). We then describe the interaction of the nuclear quantum states of the Thorium doping complex with coherent laser radiation in section~\ref{sec:equations}. In section~\ref{sec:interrogation} we propose an interrogation scheme based on counting of fluorescence photons to implement a solid-state nuclear clock and estimate the attainable fundamental limits of precision. At liquid Nitrogen temperatures, we find the clock performance to be determined by spin-spin relaxation due to interactions of the Thorium nuclear moment with neighboring nuclei. A detailed analysis of this process (presented in section~\ref{sec:numeric_performance}) for the Thorium-doped CaF$_2$ system indicates, that a fractional instability on the $10^{-19}$ level might be reached (neglecting technical limitations i.e. imposed by the interrogation laser system) within the solid-state clock approach.

\section{Thorium nuclear transition in the crystal lattice environment}
\label{sec:nucleus}

\subsection{Transition energy and isomer state lifetime}
\label{sec:energy}

The isomer transition energy of $^{229}$Th is currently derived from indirect measurements of the $\gamma$-ray spectrum resulting from the decay of $^{233}$Uranium. In 1989-1993 Helmer and Reich performed first measurements using high quality Germanium detectors (resolution from 300 to 900\,eV). They estimated the $^{229}$Th isomer state energy to be $E=3.5 \pm 1.0$\,eV~\cite{Reich90, Helmer94}. This unnaturally low value triggered a multitude of investigations, both theoretical and experimental, trying to determine the transition energy precisely, and to specify other properties of the isomer state of $^{229}$Th (such as the lifetime and the magnetic moment). However, searches for direct photon emission from the low-lying excited state have failed to report an unambiguous signal~\cite{Irwin97, Richardson98,Shaw99, Utter99}. In 2005, Guim\~araes-Filho and Helene re-analysed the old experimental data and reported the value of $E=5.5 \pm 1.0$\,eV~\cite{Filho05}. New indirect measurements with an advanced X-ray microcalorimeter (resolution from 26 to 30\,eV) were performed by Beck et al. in 2007~\cite{Beck07}. They published a new value for the transition energy $E=7.6 \pm 0.5$\,eV for the isomer nuclear transition, shifting it into the vacuum ultraviolet domain. This shift probably explains the absence of signatures of the transition in previous experiments. In 2009, Beck et al. reviewed their results taking into account the non-zero probability of the $42.43 \; \mathrm{keV} \rightarrow \, ^{229m}\mathrm{Th}$ transition (estimated as 2\,\%) and published a revised version $E=7.8 \pm 0.5$ eV~\cite{Beck09}. This value is the currently most accepted one in the community but can not be considered definite until a direct measurement is performed successfully. The dominant uncertainty in the prediction of Beck et al. is connected with the value of the branching ratio $b=1/13$~\cite{Beck07} from the 29.19\,keV level to the ground state~\cite{Sakharov10}. Estimation of this value vary in different references. In~\cite{Sakharov10} two alternative values for this branching ratio are mentioned: 25 \,\% which would result in $E=9.3 \pm 0.6$\,eV and 51\,\% which would result in $E=14.0 \pm 1.0$\,eV. In the present analysis we use the value of $7.8\pm 0.5$\,eV, scaling the obtained results to different transition energies is straightforward.

While a direct measurement of the isomer transition energy in $^{229}$Th is outstanding, the decay rate $\gamma$ of the isomer state also remains unknow. The most recent estimations of the half life of the isomer state for a \emph{bare nucleus} are based on theoretical calculations of the matrix element of the magnetic moment~\cite{Ruchowska06}. The theory was verified by comparison with experimental data for transitions at higher energies. It predicts a half-life of $T_{1/2}=(10.95 \mathrm{h})/ (0.025 E^3)$ for the isomer transition, where $E$ is given in eV. For $E=7.8$\,eV this yields $T_{1/2}=55$\,min which corresponds to a spontaneous decay rate of $\gamma_s=2.09 \cdot 10^{-4}$\,s$^{-1}$. In~\cite{Tkalya00} it is predicted that the probability of spontaneous magnetic dipole emission in a transparent non-magnetic dielectric medium with refractive index $n$ is $n^3$ times higher than in vacuum. The reason for this factor is the renormalisation of the density of emitted photon states. For a CaF$_2$ crystal, $n=1.55$ at a wavelength $\lambda=160$\,nm (corresponding to $E=7.8$\,eV), the enlargement factor is 3.7 for the spontaneous decay rate of the isomeric state. Therefore, if the crystal contains $N_{e}$ excited Thorium nuclei, the number $dN_{ph}$ of $160$\,nm photons emitted during a short time interval $dt$ can be estimated as 
\begin{equation}
dN_{ph}=N_{e}\gamma_s n^3 dt. \label{eq:1}
\end{equation}

The total decay rate $\gamma$ is determined not only by spontaneous decay of the nucleus, it can be enhanced by interactions between the nucleus and the electron shell in electronic bridge processes and bound internal conversion effects~\cite{Karpeshin07,Porsev10_3,Porsev10}. Such a decay can be accompanied by the emission of several low-energy photons with widely distributed energies~\cite{Karpeshin07}. This would make it difficult to separate decay events from optical noise sources (i.e. from scintillation or radioluminescence).

There are a number of theoretical calculations on electronic bridge processes in the Thorium system. In an isolated \emph{neutral atom} the Thorium isomer state lifetime is expected to be about 4.5\,min for $E=3.5$\,eV and about  $10^{-5}$\,s for $E=7.6$\,eV~\cite{Karpeshin07} 
\footnote{Thorium first, second, third, and fourth ionization energies are 6.31\,eV \cite{Koehler}, 11.5\,eV, 20\,eV, and 29\,eV respectively. Therefore the expected energy of the isomer transition (7.8\,eV) is above the ionization energy of the neutral Thorium atom.}. In the \emph{Th$^{3+}$ ion} for $E=7.6$\,eV the expected lifetime is the same as in a bare nucleus described above, if the valence electron is in the ground $5f_{5/2}$ state. It decreases by a factor of 20 if the valence electron is in the metastable $7s$ state~\cite{Porsev10_3}. In the \emph{Th$^{+}$ ion} the lifetime decreases $10^2-10^3$ times for $E=3.5$ and $E=5.5$\,eV~\cite{Porsev10}, there is not sufficient data about the Th$^{+}$ ion electronic structure to calculate the lifetime at $E=7.8$\,eV. Similar calculations for the \emph{Th$^{4+}$ ion}, as it will occur in the solid-state approach presented here, have yet to be performed. 

There were several experimental attempts to determine the $^{229m}$Th lifetime without knowledge of the exact isomer energy. Inamura et al. measured a temporal variation of the $\alpha$ decay activity in a sample of $^{229}$Th after excitation in a hollow-cathode electric discharge~\cite{Inamura09}. They propose a value of $T_{1/2}=2\pm 1$\,min for the half-life of the isomeric state which corresponds to a total excited state decay rate of $\gamma=6\,^{+6}_{-2}\cdot 10^{-3}$\,s$^{-1}$. Kikunaga et al. studied the $\alpha$ spectrum of chemically ``fresh" $^{229}$Th produced in the $\alpha$ decay of $^{233}$U~\cite{Kikunaga09}. The authors obtain $T_{1/2}<2$\,h with $3\sigma$ confidence. The latter experiment was not suited to measure half-lives on the order of minutes because of the long chemical preparation process. The analysis of both these experiments is significantly complicated by the presence of various chemical compositions of Thorium, mainly ThO$^+$, Th and Th$^+$ in~\cite{Inamura09} and hydroxide and chloride complexes in~\cite{Kikunaga09}. These chemical structures determine the energies of electronic levels and hence the rates of electronic bridge processes. Therefore the different results obtained in~\cite{Inamura09, Kikunaga09} may be explained by the different chemical structures and the times of preparation and measurement.

In the present work we assume the band gap of the Thorium-doped CaF$_2$ material to be sufficiently large to neglect electronic bridge processes and contributions from low-energy photons to the measured fluorescence count rates. In other words, we expect no electronic levels near 7.8\,eV in the Thorium-doped CaF$_2$ crystal. This assumption will be substantiated in a future publication~\cite{Mohn}. We perform our analysis using the theoretical value for the bare nucleus lifetime $\gamma=\gamma_s\cdot n^3=7.8\cdot 10^{-4} \, \mathrm{s}^{-1}$, enhanced by the refractive index, and assume relaxation via electronic bridge processes to be negligible.  Again, as for the energy, scaling our results to different values of the $^{229m}$Th lifetime is straightforward.

\subsection{Effects of the crystal lattice environment}
\label{sec:effects}

In this work we consider the specific system of Thorium nuclei doped into Calcium fluoride single crystals. CaF$_2$ is a standard material in UV optics, it is routinely grown and doped in many laboratories and companies worldwide using various techniques (Czochralski, Bridgman-Stockbarger, micro-pulling-down). Standard procedures for cutting, cleaving, polishing, and coating are established, CaF$_2$ is well characterized concerning radiation damage, both concerning UV laser radiation and radiation related to the nuclear decay chain of $^{229}$Th.

CaF$_2$ has a wide bandgap of about 12\,eV, values in the literature range from 11.6\,eV~\cite{BJC90} to 12.1\,eV~\cite{Rub72}. The valence band of CaF$_2$ consists of $2p$ levels of the Fluorine ions, while the bottom of the conduction band originates from the $4s$ and $3d$ orbitals of the Calcium ions. Optical transmission of better than 70\,\% in the 160\,nm wavelength regime is routinely achieved in high-purity single crystals.

The simple cubic lattice structure and the fact that only the Ca$^{2+}$ site qualifies as a doping site makes Thorium-doped CaF$_2$ amendable to full electronic structure simulations~\cite{Mohn} which is currently out of reach for significantly more complex crystal structures (i.e. LiCaAlF$_6$ as proposed in ~\cite{Rellergert10}).

Thorium atoms in a solid-state crystal lattice are confined in the Lamb-Dicke regime, the recoil energy $E_{rec}=(7.8 \, \mathrm{eV})^2/(2M_{Th}c^2)\simeq 1.4\cdot 10^{-10}\, \mathrm{eV}$ is far below the  energy required to create a phonon with a recoil momentum $p_{rec}=7.8 \, \mathrm{eV}/c$ in the lattice. Therefore the only process that could transfer the recoil energy to the lattice vibrations must include the scattering of optical phonons. However optical modes are expected to be frozen out even at room temperature~\cite{Rellergert10}. Therefore, internal and external degrees of freedom are decoupled and there is no sensitivity of the nuclear transition to recoil or first-order Doppler effects. However, the interaction between the Thorium nuclear levels and the crystal lattice environment can cause some inhomogeneous effects which will affect the performance of a solid-state nuclear clock.

We assume the $^{229}$Th ions to be chemically bound into a VUV transparent ionic crystal. Therefore the fine interaction of the Thorium nuclei with the crystal lattice environment is absent because of the absence of unpaired electronic spins. The remaining interaction is the hyperfine interaction where the Hamiltonian $\hat{H}'=\hat{H}_{HFS}$ can be represented in a multipole expansion $\hat{H}_{HFS}=H_{E0}+H_{M1}+H_{E2}+...$~\cite{Rellergert10}. In the following we will discuss the individual terms in more detail.

The \emph{electric monopole shift} ($H_{E0}$) arises from contact interaction between the electron cloud and the finite nuclear volume. As the ground state and the excited nuclear state can have different sizes, this effect leads to a shift of the nuclear transition relative to a bare nucleus. Typical monopole shifts are about $1$\,GHz~\cite{BerengutIsomer09}, they will depend on the specific choice of the host crystal. However, this effect will be identical (up to temperature effects) for all nuclei having the same position within the lattice and hence deteriorate the solid-state nuclear clock accuracy, but not the stability. 
Temperature effects appear because the electron density at a given lattice site is temperature dependent. This temperature dependence is estimated as 10\,kHz/K~\cite{Rellergert10} which leads to a broadening of about $10$\,Hz for a temperature stability of $\Delta T \approx 1$\,mK which is attainable by standard techniques. It should be noted that a certain contribution to the decoherence rate can be caused by crystal lattice vibrations. These vibrations will modulate the electronic density in the vicinity of the nucleus thus broaden the transition line. A detailed study of this broadening effect is beyond the scope of this article. We expect that this effect, if relevant at all, can be effectively suppressed by cooling of the crystal lattice.

\begin{figure}
\begin{center}
\resizebox{0.3\textwidth}{!}{ \includegraphics{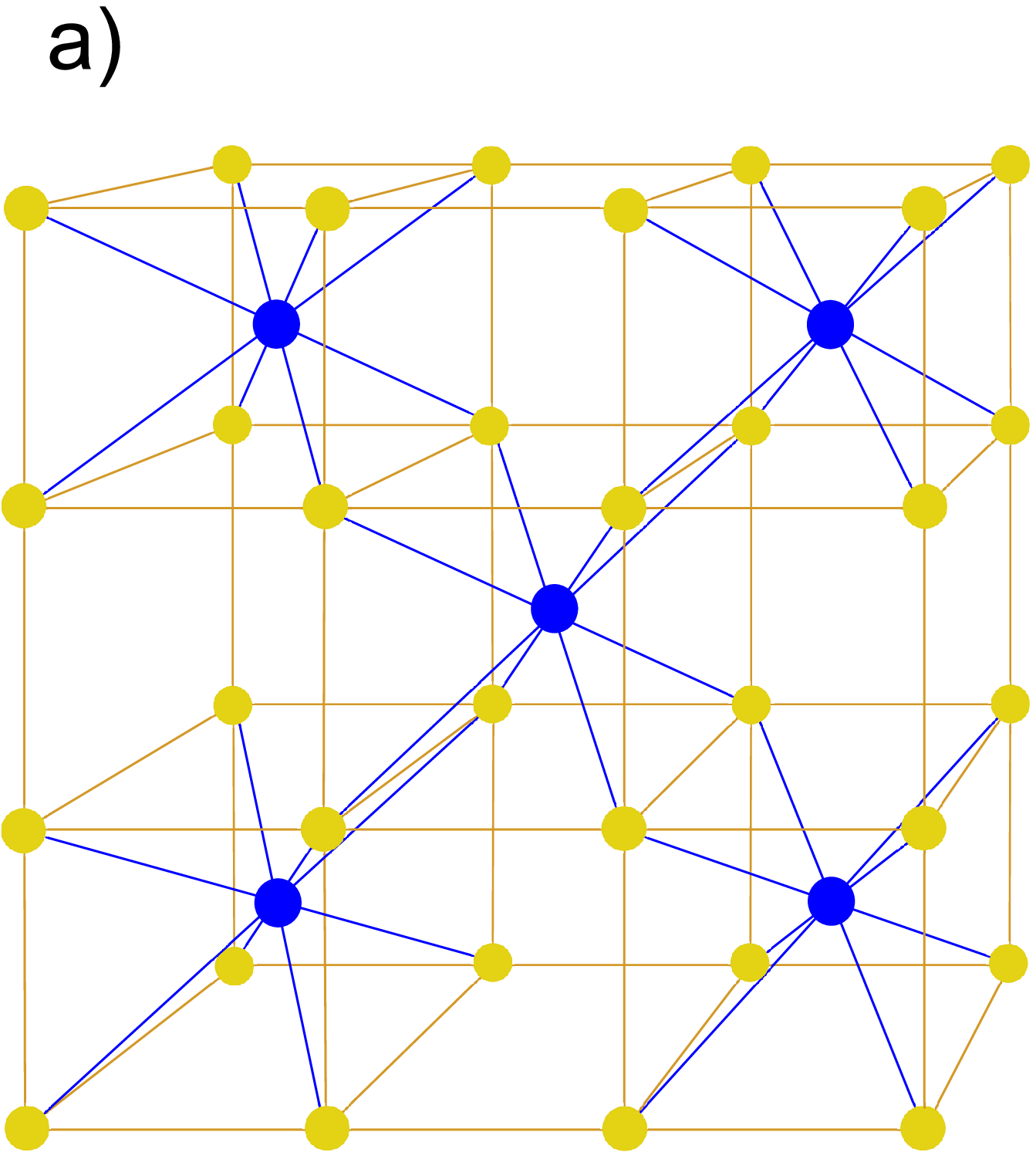}}
\hspace*{7mm}
\resizebox{0.3\textwidth}{!} { \includegraphics{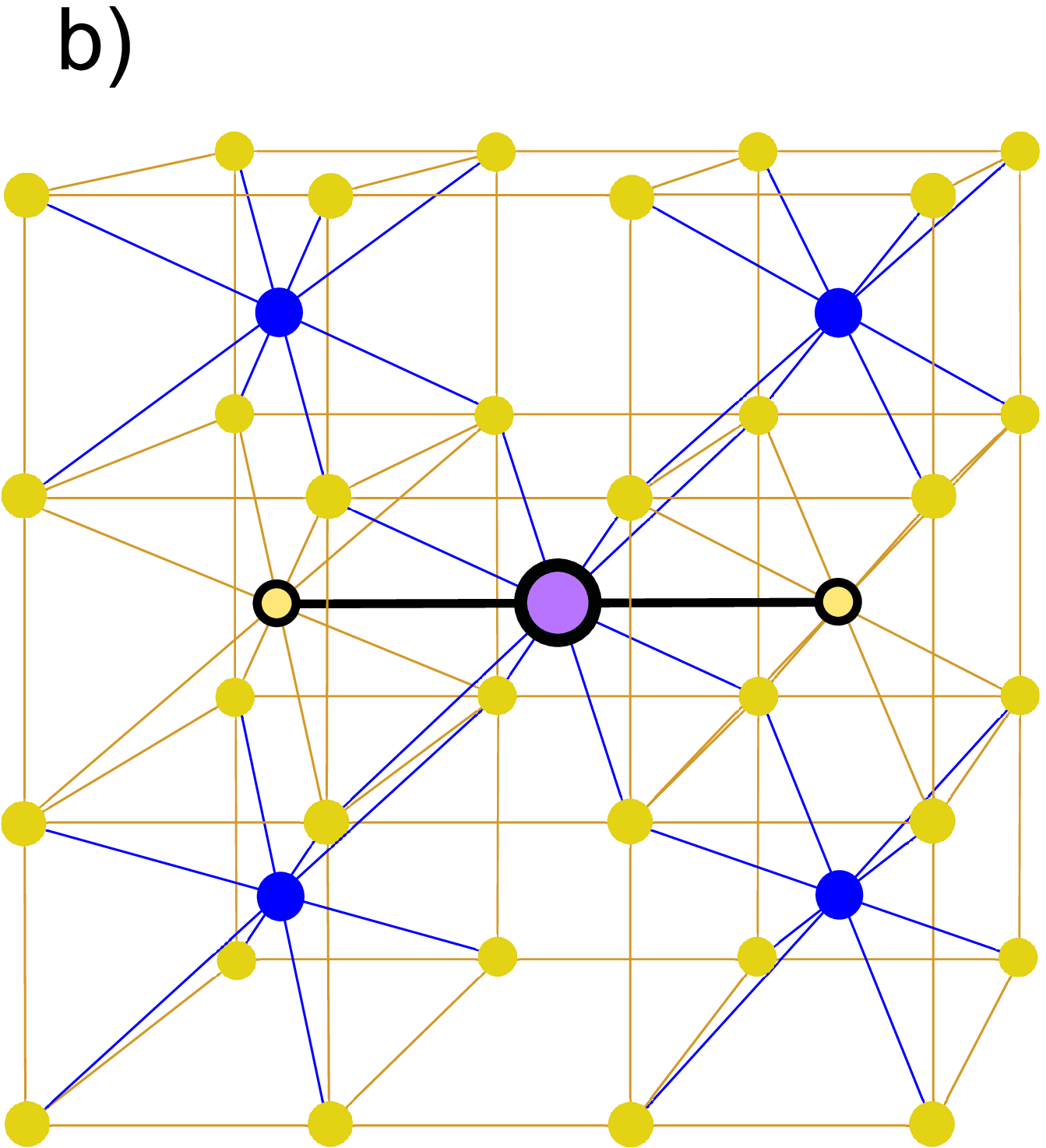}}
\hspace*{7mm}
\resizebox{0.13\textwidth}{!} { \includegraphics{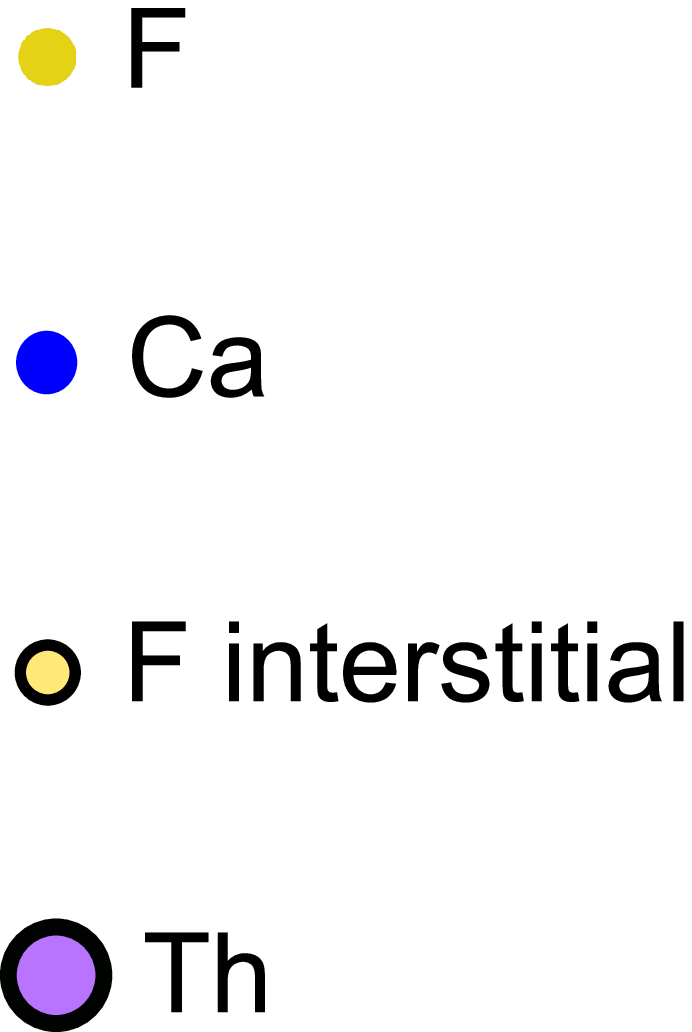}}
\end{center}
\def\baselinestretch{1.1}
\caption{CaF$_2$ crystal lattice (a) and Thorium inclusion with charge compensation by two in-line Fluorine interstitials (b).}
\def\baselinestretch{1.5}
\label{fig:f1}
\end{figure}

The \emph{electric quadrupole shift} ($H_{E2}$) appears due to the electric quadrupole moment $Q$ of the $^{229}$Th nucleus in the presence of a non-vanishing electric field gradient at the position of the nucleus. Reported experimental values for the quadrupole moment in the ground state are $Q_g=3.149(32)$\,eb~\cite{Bemis88} and $Q_g=3.11(16)$\,eb~\cite{Campbell11PRL}, the value $Q_{is}$ of the quadrupole moment in the isomer state is still unknown, estimations~\cite{Tkalya11} give $Q_{is}\approx 1.8$\,eb. If the electric field gradient in the principal axes has two equal components, $\phi_{xx}=\phi_{yy}=-\phi_{zz}/2$, the matrix element of the quadrupole interaction term can be written as~\cite{Greenwood}
\begin{equation}
\langle Im|\hat{H}_{E2}|Im'\rangle =\delta_{mm'}\frac{Q \phi_{zz}}{4I (2I-1)}\left[3m^2-I(I+1)\right]
\label{eq:2}
\end{equation}
where the $z$ axis is the axis of symmetry of the electric field, $\phi_{zz}=\partial^2\varphi/\partial z^2$, $I$ and $m$ are the nuclear angular moment and its projection on the axis of symmetry of the electric field respectively. In a Th-doped CaF$_2$ crystal, the Thorium ion will most probably replace a Ca$^{2+}$ ion in a Th$^{4+}$ state. To recover an ionic (and hence transparent) crystal, the two remaining electrons will be localized by a charge compensation mechanism, either by introducing two additional Fluorine (F$^-$) interstitials or by creating a vacancy by leaving a neighboring Ca$^{2+}$ site empty. First simulations indicate a preference for charge compensation by interstitials which will be the main contribution to the electric field gradient~\cite{Jackson}.
 
\begin{figure}
\begin{center}
\resizebox{0.5\textwidth}{!}
{ \includegraphics{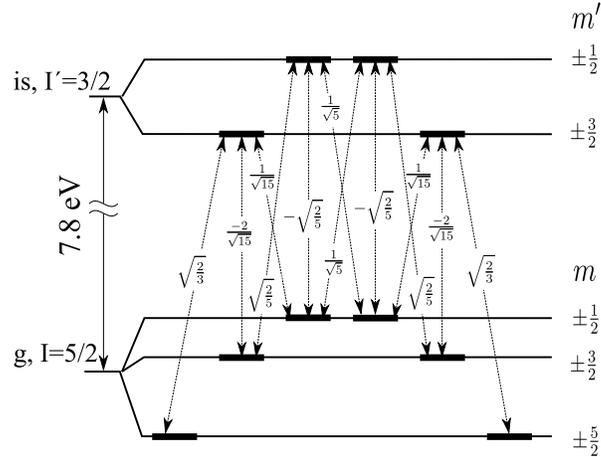}}
\end{center}
\def\baselinestretch{1.1}
\caption{Quadrupole structure of the ground and isomer state of the $^{229}$Th nucleus in the CaF$_2$ crystal lattice with charge compensation as presented in Figure~\ref{fig:f1}(b). Light arrows represent the possible magnetic dipole transitions, matrix elements are proportional to the indicated Clebsch-Gordan coefficients $C_{Im1q}^{I'm'}\, (q=m'-m)$. The quantization axis $z$ is chosen along the axis of symmetry of the electric field gradient.}
\def\baselinestretch{1.5}
\label{fig:scheme}
\end{figure}

Consider the charge compensation by two F$^{-}$ interstitials as shown in Figure~\ref{fig:f1}\,(b). We expect the geometrical lattice distortion around the dopant to be negligible. Using the Ca$^{2+}$ lattice constant $a=5.6$ \AA \, and the quadrupole antishielding factor $\gamma_{\infty}=-177.5$~\cite{Feiock69} we estimate $\phi_{zz} \approx -5.1 \cdot 10^{18} \, \mathrm{V/cm^2}$.
This leads to quadrupole splittings of the order of a few hundred MHz which can be easily resolved in laser spectroscopy (see Figure~\ref{fig:scheme}). Note that this splitting could also be directly measurend experimentally using standard NMR techniques, which would give valuable information on the details of the charge compensation mechanism.

The quadrupole interaction of the nucleus with phononic modes of the crystal lattice leads to inhomogeneous broadening and mixing between different quadrupole sublevels as the the typical distances between the quadrupol sublevels are much smaller than characteristic phonon energies. Comparison with experimental data for the quadrupole spin-lattice relaxation of the Halogen nuclei in CsBr and CsI crystals~\cite{Wikner60} and scaling the theory of Van Kranendonk~\cite{VanKranendonk54} and Yoshida and Moriya~\cite{YoMo56} to the CaF$_2$ crystal yields a relaxation and mixing rate of the order of $2 \pi \cdot (5 - 100)$ Hz at room temperature. This rate scales as $T^{2}$, leading to about $2 \pi \cdot (0.3 - 7)$ Hz at liquid Nitrogen temperature. The temperature of the crystal should however not be so low to entirely suppress the mixing, as populations accumulating in quadrupole sublevels not interacting with the laser field could lead to a loss of signal or neccessitate additional repump lasers.

The \emph{2$\mathrm{^{nd}}$ order Doppler effect} leads to a shift and broadening of the isomer transition due to vibrations of the nuclei. Using a classical harmonic oscillator model, we find the shift $S=-\omega \langle v^2 \rangle /(2 c^2)=-3 \omega k_B T/(2M_{Th}c^2)$ and the broadening $\Delta \omega=\omega \langle \Delta v^2 \rangle /(2 c^2) = \sqrt{6}\,\omega k_B T/(2M_{Th}c^2)$. Here $\omega$ is the isomer transition frequency, $v$ the velocity of the Thorium nuclei, $\langle \Delta v^2 \rangle=\sqrt{\langle v^4 \rangle - \langle v^2 \rangle^2}$. This leads to a shift of about $2\pi \cdot 1.14 \cdot T$\, (in Hz), and to a broadening of about $2\pi \cdot 0.93 \cdot T$\, (in Hz) where $T$ is the temperature in Kelvin. At liquid Nitrogen temperature we expect a shift of about $2 \pi \cdot 88$\, Hz and a broadening of about $2 \pi \cdot 72$\, Hz.

The term $H_{M1}$, i.e. the \emph{magnetic dipole interaction} with randomly oriented spins of surrounding nuclei also contributes to broadening. The characteristic energies of spin-spin interactions correspond to the temperature range of the order 1 microKelvin, the broadening is expected to be temperature-independent for experimentally relevant conditions, in contrast to all other broadenings discussed above, which decrease with temperature. A detailed evaluation (see Appendix A) yields a contribution to the optical relaxation rate from $\sim2\pi\cdot 84$\,Hz for the $|I=5/2,m=\pm 3/2\rangle \leftrightarrow |I'=3/2,m'=\pm 1/2\rangle$ optical transition to $\sim2\pi\cdot 251$\,Hz for the $|I=5/2,m=\pm 5/2\rangle \leftrightarrow |I'=3/2,m'=\pm 3/2\rangle$ optical transition.

To resume, we expect a global uniform shift of about 1 GHz dominated by the electric monopole shift compared to the bare nucleaus isomer transition frequency (assuming identical doping complexes and charge compensation). Main sources of inhomogeneous broadening are magnetic dipole interaction and 2$^{nd}$ order Doppler effect, leading to an ensemble linewidth of $2\pi\cdot(84-251)$\,Hz and $2\pi \cdot 0.93$\,Hz$ \,\cdot \,T$ respectively where $T$ is the temperature in Kelvin. Taking the ensemble linewidth of $2\pi\cdot 150$\,Hz we expect a quality factor $Q=1.26\cdot 10^{13}$ for the solid state clock transition. Note that crystal field shifts and broadenings (and their temperature dependence) can directly be studied with narrow band lasers, transferring Mößbauer spectroscopy into the optical regime.
\section{Excitation of Thorium nuclei by laser radiation}
\label{sec:equations}

Let us consider an ensemble of $^{229}$Th$^{4+}$ ions implanted into a CaF$_2$ crystal and interacting with a narrow-band vacuum ultraviolet laser whose frequency $\omega_L$ is close to the frequency $\omega$ of the isomer transition. Here we consider only the ions in one charge-compensating configuration, i.e. the one presented in Figure \ref{fig:f1}\,(b). All other configurations will not interact with the laser field due to different quadrupole shifts.
Th$^{4+}$ ions are iso-electronic with the noble gas Radon with zero angular momentum and zero spin due to the electronic shell. In the following we neglect the electronic shell structure as well as higher excitation levels of the Th nucleus. The nuclear ground and isomer states are degenerate with respect to the sign of projections $m$ and $m'$ and with respect to angular momenta $I$ and $I'$ up to interactions with randomly oriented spins of surrounding nuclei. The magnetic field $\vec{B}$ of the spectroscopy laser radiation can be written as $\vec{B}=\vec{B}_0 \, \cos \omega_L t$. 

The Hamiltonian $\hat{H}$ of the Thorium nucleus in resonance approximation can be represented as
\begin{equation}
\hat{H}=\hat{H}_0 + \hbar \sum_{e,g} V_{eg}\left[|g\rangle\langle e|e^{i\omega_L t}+
|e\rangle\langle g|e^{-i\omega_L t} \right]+\hat{H}'. \label{eq:3}
\end{equation}
Here, the indices $e$ and $g$ indicate the sublevels of the nuclear excited and ground state respectively, $V_{eg}=B_{0q} \mu_{eg}^q/(2\hbar)$, $B_{0q}$ is the $q$-th covariant component of $\vec{B}_0$,
$\mu_{eg}^q$ is the $q$-th contravariant component of the matrix element $\vec{\mu}_{eg}$ of the nucleus magnetic momentum operator in cyclic coordinates, $q=m'-m$, $\hat{H}_0$ is the Hamiltonian of the free Thorium ion. The term $\hat{H}'$ describes the interactions of the nucleus with the crystal lattice leading to shifts (which could also be added to $\hat{H}_0$) and relaxation terms which will be discussed later. The set of equations for the density matrix $\rho_{ij}$ with relaxation terms in the rotating frame reads:
\begin{eqnarray}
\dot{\rho}_{ee}&=&-i\sum_g V_{eg} \left[\rho_{ge}-\rho_{eg}\right]-\gamma \rho_{ee}+\sum_{e'}\Gamma_{||}^{ee'}\rho_{e'e'}; \label{eq:4}\\
\dot{\rho}_{ee'}&=&-i\sum_g \left[V_{eg} \rho_{ge'}-V_{e'g} \rho_{eg}\right]-(i\omega_{ee'}+\Gamma_\perp^e) \rho_{ee'},  \quad e\neq e';  \label{eq:5}\\
\dot{\rho}_{eg}&=&\Big[i(\omega_L-\omega_{eg})-\Gamma'\Big]\rho_{eg}-i\sum_{g'}V_{eg'} \rho_{g'g}+i\sum_{e'}V_{e'g}\rho_{ee'}; \label{eq:6}\\
\dot{\rho}_{gg}&=&-i\sum_e V_{eg} \left[\rho_{eg}-\rho_{ge}\right]+\sum_{e}\gamma_{eg}\rho_{ee}+ \sum_{g'}\Gamma_{||}^{gg'}\rho_{g'g'}; \label{eq:7}\\
\dot{\rho}_{gg'}&=&-i\sum_e \left[V_{eg} \rho_{eg'}-V_{eg'} \rho_{ge}\right]-(i\omega_{gg'}+\Gamma_\perp^g) \rho_{gg'}, \quad g\neq g'.\label{eq:8}
\end{eqnarray}
Here $\Gamma_{\perp}^{e,g}$ and $\Gamma_{||}^{i,j}$ are decoherence and mixing rates between magnetic states of the same quadrupole energy level, $\Gamma'$ is the relaxation rate of optical coherence $\rho_{eg}$, $\omega_{ij}$ is the difference between the frequencies of the $i$-th and $j$-th level, $\gamma_{eg}=\gamma\cdot |C_{Im1q}^{I'm'}|^2$ is the rate of spontaneous $|e\rangle \rightarrow |g\rangle$ transitions.

In this work we suppose that the spectroscopy laser field is weak, i.e. 
\begin{equation}
V_{eg}\ll \min[\Gamma', \Gamma_{\perp}^{e,g}, \Gamma_{||}^{ee'}, \Gamma_{||}^{gg'}]. \label{uneq:9}
\end{equation}
Then, from equations (\ref{eq:4}) -- (\ref{eq:8}) and from the normalization condition $\sum_i \rho_{ii}=1$ we derive:
\begin{equation}
\rho_{gg'}=\frac{\delta_{gg'} (1-\rho_{exc})}{2I+1}\, , \quad \rho_{ee'} = \delta_{ee'}\frac{\rho_{exc}}{2I'+1}\, , \label{eq:10}
\end{equation}
where $\rho_{exc}=\sum_{e}\rho_{ee}$ is the total excited state population. From equations (\ref{eq:6}), (\ref{eq:8}), and (\ref{eq:10}) we obtain:
\begin{equation}
\dot{\rho}_{exc}= \sum_{e,g}\frac{2 \Gamma' V_{eg}^2}{\Gamma'^2+(\omega_L-\omega_{eg})^2}
\left(\frac{1}{6}-\frac{5 \rho_{exc}}{12}\right) - \gamma \rho_{exc}. \label{eq:11}
\end{equation}
The quadrupole splitting in the ground and the isomer state is of the order of hundred MHz which is much greater than the expected value of $\Gamma'$ (about a few kHz). Hence only transitions between certain quadrupole sublevels of the ground and excited states should be taken into account. Let us consider the case where then laser is tuned close to the transition between $|g_{1,2}\rangle=|I=5/2,m=\pm 3/2\rangle$ and $|e_{1,2}\rangle=|I'=3/2,m=\pm 1/2\rangle$.  This is the transition with the lowest relaxation rate of optical coherence, see Appendix~A. In the following we will name this transition the \emph{clock transition} and use $\omega_{eg}=\omega$. We introduce the optical detuning $\Delta=\omega_L-\omega$. Suppose, for the sake of simplicity, that the radiation has a certain polarization along $q'$-th cyclic unit vector, so $B_{0q}=|\vec{B}_{0}|\delta_{qq'}$.
In dipole approximation, $\gamma_{s} = \frac{4 \omega^3}{3 \hbar c^3}\sum_g|\mu^q_{eg}|^2$. According to the Wigner-Eckart theorem, 
$\mu_{eg}^q=||\mu||C_{Im1q}^{I'm'}/\sqrt{2I'+1}$, where $||\mu||=\sqrt{\frac{4\pi}{3}B(M1)(2I'+1)}=0.65 \, \mu_N$ is a reduced dipole moment ($B(M1)=0.025 \, \mu_N^2$ is the reduced transition probability \cite{Ruchowska06}), $C_{Im1q}^{I'm'}$ is a
Clebsch-Gordan coefficient, $c$ is the speed of light in vacuum. 
Using $\sum_{m,q}\left|C_{Im1q}^{I'm'}\right|^2=1$, we express
\begin{equation}
||\mu||=\sqrt{\frac{3\hbar c^3 \gamma_s(2 I'+1)}{4 \omega^3}}\, ,\label{eq:12}
\end{equation}
and therefore
\begin{equation}
V_{eg}^2=\frac{B^2_{0q} ||\mu||^2}{4 \hbar^2 (2 I'+1)} \left|C_{Im1q}^{I'm'}\right|^2 \, .\label{eq:13}
\end{equation}
Averaging over possible spatial orientations of the electric field gradient with respect to the polarization of the laser, and expressing the amplitude $|\vec{B}_0|$ via the intensity $I_0$ of the resonance radiation: $|\vec{B}_0|^2=8 \pi I_0/c$, we finally obtain
\begin{equation}
\dot{\rho}_{exc}= \frac{R}{1+\Delta^2/\Gamma'\,^2} - \left[\gamma+\frac{5}{2}\cdot\frac{R}{1+\Delta^2/\Gamma'\,^2}\right] \rho_{exc}\, , \label{e:15}
\end{equation}
where the excitation rate $R$ is
\begin{equation}
R=\frac{\langle V_{e_1g_1}^2+V_{e_2g_2}^2 \rangle}{3}=\frac{2 \pi}{15} \; \frac{c^2 I_0}{\hbar \omega^3} 
\frac{\gamma_s}{\Gamma'}\, .\label{e:16}
\end{equation}
Here we denote the averaging over spatial orientations by angled brackets and we have taken into account the values of the $^{229}$Thorium nuclear angular momenta $I=5/2$ and $I'=3/2$ in the ground and excited state respectively.

\section{Interrogation scheme and fundamental limits of stability}
\label{sec:interrogation}

\begin{figure}
\begin{center}
\resizebox{0.4\textwidth}{!}
{ \includegraphics{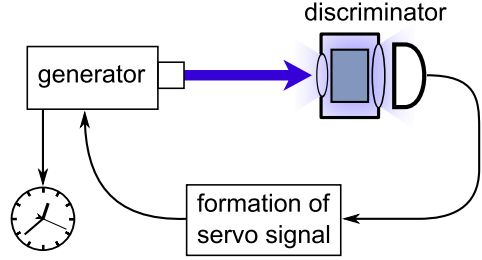}}
\end{center}
\def\baselinestretch{1.1}
\caption{Schematic representation of a passive quantum frequency standard constructed from a generator (laser), a quantum discriminator, and a feedback loop. Linking the stabilized generator to an electronic counting system (i.e. through a frequency comb system) allows the realization of a clock.}
\def\baselinestretch{1.5}
\label{fig:f2}
\end{figure}

Any passive quantum frequency standard consists on three main elements: a generator producing some periodic signal (a laser in the case of an optical frequency standard) with good short-time stability and a frequency that can be tuned around a ``nominal" frequency $\omega_n$, a quantum discriminator, and a servo loop determining and correcting the offset $f=\omega_n-\omega$ between the nominal frequency $\omega_n$ and the frequency $\omega$ of the clock transition according to the result of the interrogation of the quantum discriminator. If we add a system to count the number of oscillations of the generator, we obtain a clock (see Figure~\ref{fig:f2}).

Usually, the Rabi, Ramsey~\cite{Riehle}, or hyper-Ramsey~\cite{Yudin10} interrogation scheme is used in high-performance frequency standards. However, all these methods require the exact knowledge of the quantum state of the quantum discriminator at the beginning of any interrogation cycle. If the quantum discriminator consists of one or a small number of ions it is possible to know the quantum state precisely at the end of any interrogation cycle (and therefore at the beginning of any next cycle) as the interrogation measurement leads to a projection of the quantum state onto a well-defined basis. If the quantum discriminator consists of a large number of trapped ions it is possible either to reload the trap with ions of a known quantum state or to repump the ions back to the initial quantum state via some short-living state. Note that for Thorium ions in an ion trap such "repumping" of the isomer state could be performed using an excitation of the electronic shell to a resonant metastable state which will lead to an amplification of electronic bridge processes~\cite{Porsev10_3}. 

In the solid-state nuclear frequency standard such a repumping seems to be impossible due to the absence of an accessible metastable state in the band gap. Another fundamental problem arised from the huge difference between the relaxation time of the excited state population (presumably tens of minutes) and of the short coherence time between the ground and excited state (milliseconds) due to crystal lattice effects. At least one Rabi or Ramsey interrogation cycle should be completed during the time interval $t < 1/\Gamma'$ whereas the time $t_{p}$ needed for bringing the nuclei back to the ground state should be $t_p > 1/\gamma$. Therefore the Rabi or Ramsey interrogation time would take only a tiny ($\approx10^{-6}$) fraction of the total interrogation cycle which would significantly reduce the clock performance due to the Dick effect~\cite{Dick}. Moreover, Ramsey or Rabi interrogation schemes require quite strong laser fields, able to excite a significant number of nuclei during a short interrogation time.
\subsection{Clock interrogation by fluorescence spectroscopy}

\begin{figure}[t]
\begin{center}
\resizebox{0.5\textwidth}{!}
{ \includegraphics{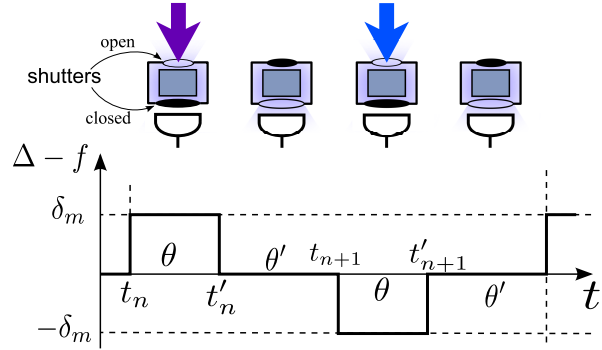}}
\end{center}
\def\baselinestretch{1.1}
\caption{Interrogation cycle of quantum discriminator as discussed in the text. Note that this cycle is opperated continuously without any deadtime.}
\def\baselinestretch{1.5}
\label{fig:f3}
\end{figure}

To tackle these problems we propose an interrogation scheme based on counting of spontaneous nuclear decay fluorescence photons after illumination of the quantum discriminator by the spectroscopy laser. The $(n,\, n+1)$ interrogation cycle consists of four time intervals, as depicted in Figure~\ref{fig:f3}. In the first interval $\theta$ between $t_n$ and $t'_n$ the crystal sample is illuminated by laser radiation with the frequency detuned by $\delta_m$ to the blue side of the nominal frequency $\omega_n$ of the generator. Then a laser shutter is closed and the fluorescence photons are counted by a photodetector during the second time interval $\theta'$ between $t'_n$ and $t_{n+1}$. Then these two operation are repeated, but with the laser frequency shifted to the red side by $\delta_m$ of the nominal value. After the last measurement we determine the frequency offset $f=\omega_n-\omega$ and make a correction to the nominal frequency. As no further state initialization, repumping, or sample preparation is required, the next interrogation cycle can start imediately.

We will now consider a single interrogation cycle in more detail. Solving equation~(\ref{e:15}) we express the excited state population $\rho_{exc}$ at time $t'_n$ as
%
\begin{eqnarray} 
\rho_{exc}(t'_n)&=&\rho_{exc}(t_n)\exp\left[-\theta\cdot\left(\gamma+\frac{5R/2}{1+(\delta_m+f)^2/\Gamma'\,^2}\right)\right]\nonumber \\
&&+\frac{R}{\gamma[1+(\delta_m+f)^2/\Gamma'\,^2]+5R/2}  \label{e:17} \\
&&\times \left\{1- \exp\left[-\theta\cdot\left(\gamma+
\frac{5R/2}{1+(\delta_m+f)^2/\Gamma'\,^2}\right)\right] \right\}. \nonumber
\end{eqnarray}
%
The number $N_n$ of nuclear fluorescence photons counted by the photodetector during time $\theta'$ between $t'_n$ and $t_{n+1}$ is
\begin{equation}
N_{n}=N_{eff}\cdot \rho_{exc}(t'_n)\cdot(1-e^{-\gamma \theta'}), \label{e:18}
\end{equation}
where 
\begin{equation}
N_{eff}=N_{Th}\cdot \frac{\gamma_s n^3}{\gamma}\cdot\frac{k\Omega}{4\pi}  \label{e:19}
\end{equation}
is the product of the number $N_{Th}$ of interrogated Thorium nuclei, the ratio $\gamma_s n^3/\gamma$ of the rate of radiative decay of the isomer state to the total decay rate enhanced by possible electronic bridge processes, and the probability to detect the emitted photon. Here $k$ is the quantum efficiency of the photodetector and $\Omega$ is the effective solid angle covered by the detector.

In turn, the excited state population at time $t_{n+1}$ is equal to
\begin{equation}
\rho_{exc}(t_{n+1})=\rho_{exc}(t'_{n})\,e^{-\gamma \theta'}\, , \label{e:20}
\end{equation}
at time $t'_{n+1}$ it is equal to
\begin{eqnarray}
\rho_{exc}(t'_{n+1})&=&\rho_{exc}(t_{n+1})\exp\left[-\theta\cdot\left(\gamma+\frac{5R/2}{1+(f-\delta_m)^2/\Gamma'\,^2}\right)\right]\nonumber \\
&&+\frac{R}{\gamma[1+(f-\delta_m)^2/\Gamma'\,^2]+5R/2}  \label{e:21} \\
&&\times \left\{1- \exp\left[-\theta\cdot\left(\gamma+
\frac{5R/2}{1+(f-\delta_m)^2/\Gamma'\,^2}\right)\right] \right\}, \nonumber
\end{eqnarray}
and in turn
\begin{equation}
N_{n+1}=N_{eff}\cdot \rho_e(t'_{n+1})\cdot(1-e^{-\gamma \theta'}). \label{e:22}
\end{equation}

Let us introduce the ``evolution rate in the presence of laser field''
\begin{eqnarray}
G(\Delta)&=&\gamma+\frac{5R/2}{1+\Delta^2/\Gamma'^2}, \label{e:23}
\end{eqnarray}
and the functions
\begin{eqnarray}
a(\Delta)&=&\frac{2 N_{eff}}{5}\cdot \left(1-e^{-\gamma \theta'} \right)\cdot
\left(1-\frac{\gamma}{G(\Delta)}\right)\cdot \left(1-e^{-G(\Delta) \theta} \right), \label{e:24} \\
b(\Delta)&=&e^{-G(\Delta) \theta-\gamma \theta'}. \label{e:25}
\end{eqnarray}
We then can rewrite expressions (\ref{e:18}) and (\ref{e:22}) as:
\begin{equation}
N_{n}=a(f+\delta_m)+b(f+\delta_m)N_{n-1},  \label{e:26}
\end{equation}
\begin{equation}
N_{n+1}=a(f-\delta_m)+b(f-\delta_m)N_{n},  \label{e:27}
\end{equation}
where $N_{n-1}$ is the number of photon counts collected in the last measurement of the previous interrogation cycle, i.e. between $t'_{n-1}$ and $t_n$.

We suppose that the generator offset $f$ is small, i.e., $f \ll \Gamma'$. We then can expand the functions $a$ and $b$ near the working point $\delta_m$ into a Taylor series:
\begin{equation}
a(f\pm \delta_m)=a_0\mp a_1 f, \quad b(f\pm \delta_m)=b_0\pm b_1 f. \label{e:28}
\end{equation}
Using equations (\ref{e:26}), (\ref{e:27}) and (\ref{e:28}), we express the frequency offset $f$ as
\begin{equation}
f=\frac{N_{n+1}-N_n-b_0(N_n-N_{n-1})}{2a_1-b_1(N_n-N_{n-1})}. \label{e:29}
\end{equation}

Now let us calculate the error $\delta f$ in determining the frequency offset $f$. Here we suppose that $N_{n+1}$ and $N_n$ are Poissonian random numbers with mean values described  by (\ref{e:26}) and (\ref{e:27}), i.e. the error of $N_i$ is $\sqrt{\overline{N}_i}$. The average value $\overline{N_{i}}$ is
\begin{equation}
\overline{N}_{i}\simeq a_0+b_0 \overline{N}_{i}\simeq \frac{a_0}{1-b_0}. \label{e:30}
\end{equation}
From (\ref{e:26}) --- (\ref{e:30}) we express the error of the frequency offset as
\begin{equation}
\delta f=\frac{\sqrt{a_0(1-b_0)(1+b_0+b_0^2)}}{\sqrt{2}\left[a_1(1-b_0)-b_1a_0\right]}. \label{e:31}
\end{equation}
Substituting explicit expressions for $a_0$, $b_0$, $a_1$, and $b_1$ into (\ref{e:31}), we obtain
\begin{eqnarray}
\delta f &=&  \frac{(\Gamma'+\delta_m^2/\Gamma')^2\sqrt{1-e^{-G \theta}}}{2 R \delta_m \sqrt{5 N_{eff}(1-e^{-\gamma \theta'})}} \nonumber \\
&\times & \frac{\sqrt{\left(1-\frac{\gamma}{G} \right)(1-e^{-G\theta-\gamma \theta'})(1+e^{-G \theta-\gamma \theta'}+e^{-2 G \theta-2 \gamma \theta'})}}
{\frac{\gamma}{G^2}(1-e^{-G \theta})(1-e^{-G \theta - \gamma \theta'})+\left(1-\frac{\gamma}{G}\right)\theta e^{-G\theta}(1-e^{-\gamma \theta'})}\, ,  \label{eq:32}
\end{eqnarray}
where $G=G(\delta_m)$.

Expression~(\ref{eq:32}) represents a fundamental lower limit on the error of the frequency offset for one interrogation cycle as depicted in Figure~\ref{fig:f3}. If the generator has good long-term stability, and the algorithm for frequency stabilization succeeds in the total correction of the measured offset (\ref{e:29}), then expression (\ref{eq:32}) is the {\em frequency error} corresponding to an \emph{absolute frequency instability} of the solid-state nuclear clock after one interrogation cycle. 

The frequency error (\ref{eq:32}) depends on the specific properties of the quantum discriminator ($N_{Th}$, $\gamma$, $\Gamma'$), the detection system ($k\, \Omega$), on the intensity of laser radiation determining the excitation rate $R$, and on the regime of interrogation ($\theta$, $\theta'$, and $\delta_m$). In practice the total interrogation time $t=2(\theta+\theta')$ will be imposed by the short-term stability of the interrogation laser (i.e. the duration over which frequency drifts and fluctuations are smaller than $\delta f$). Also the laser intensity (and therefore the excitation rate $R$) will have a technical upper limit. On the other hand, the working point $\delta_m$ and the relative irradiation time 
\begin{equation}
t_R=\frac{\theta}{\theta+\theta'}.\label{eq:33}
\end{equation}
can be adjusted freely. It is therefore interesting to determine the optimal values of $t_R$ and $\delta_m$ and study the dependence of the fractional frequency error $\delta f/\omega$ as a function of total interrogation time $t$ and excitation rate $R$ for optimized $t_R$ and $\delta_m$. Because the general expression~(\ref{eq:32}) seems to be too complex for a direct interpretation, we will first focus on limiting cases of short and long interrogation times without restricting to explicit experimental parameters. In section~\ref{sec:numeric_performance} we will present numerically optimized results for the specific system of Thorium-doped Calcium fluoride crystals. A comparison between the analytic expressions and numerical optimizations can be found in Appendix~B.

\subsection{Frequency error for short interrogation time}

Short interrogation time $t=2(\theta+\theta')$ refers to $\gamma \theta' \ll 1$ and $G \theta \ll 1$, so the irradiation time and the detection time are shorter than the overall decay time and the inverse excitation rate respectively. In this case we can expand the exponents in (\ref{eq:32}) into a Taylor series and keep only the main terms. We obtain
\begin{equation}
\delta f =  \frac{(\Gamma'+\delta_m^2/\Gamma')^2}{2 R \delta_m \sqrt{5 N_{eff}}}
\cdot \frac{\sqrt{3 \left(G-\gamma \right) (G\theta+\gamma\theta')} }{(\theta+\theta')\sqrt{\gamma^3\theta\theta'}}, \label{eq:34}
\end{equation}
the absolute frequency error scales as $t^{-3/2}$ with the total duration of the interrogation cycle. First we fix $\delta_m$ and $t$ and find the optimal value of $t_R$. Therefore we should minimize the function
\begin{equation*}
\frac{Gt_R+\gamma(1-t_R)}{t_R(1-t_R)}
\end{equation*}
producing the optimal value for $t_R$:
\begin{equation}
t_R=\frac{\sqrt{G\gamma}-\gamma}{G-\gamma}. \label{eq:35}
\end{equation}

The frequency error $\delta f$ for the optimized value of $t_R$ is
\begin{equation}
\delta f = \frac{\Gamma'(1+\delta_m^2/\Gamma'^2)^2}{ R (\delta_m/\Gamma') \sqrt{5 N_{eff}}}
\cdot \frac{ \sqrt{6 \left(G(\delta_m)-\gamma\right)^3}\left(\sqrt{G(\delta_m)}-\sqrt{\gamma}\right)}
{\sqrt{\gamma^3t^3}}. \label{eq:36}
\end{equation}
Taking into account (\ref{e:23}) and minimizing (\ref{eq:36}) we can find optimal values for the working point $\delta_m$ and the frequency error $\delta f$. 

In the limit of weak field, i.e., for $R\ll \gamma$, we find
\begin{equation}
\delta_m=\frac{\Gamma'}{\sqrt{2}}, \quad \mathrm{and}  \label{eq:37}
\end{equation}
\begin{equation}
\delta f =\frac{9 \Gamma'}{\gamma \, t^{3/2} \sqrt{R N_{eff}}}. \label{eq:38}
\end{equation}

In the limit of strong field, i.e., if $R\gg \gamma$, we find
\begin{equation}
\delta_m=\Gamma',  \quad \mathrm{and}   \label{eq:39}
\end{equation}
\begin{equation}
\delta f =\sqrt{\frac{30}{N_{eff}}}\,\frac{\Gamma'}{\gamma^{3/2}\, t^{3/2}}\, . \label{eq:40}
\end{equation}

\subsection{Frequency error for long interrogation time}
We now consider the case where the interrogation time is long compared to the total decay time, i.e. when  $\gamma t \gg 1$. Here we can not neglect the exponents in (\ref{eq:32}) because the optimal value either of $\theta$ or of $\theta'$ can be comparable to $1/\gamma$ or $1/G$ respectively. From the structure of (\ref{eq:32}) we expect a small optimal value of $\theta$ because of the presence of $\sqrt{1-e^{-G\theta}}$ in the numenator. Therefore we suppose that $\gamma \theta' \gg 1$ and neglect the terms in $e^{-\gamma \theta'}$. Under these asumptions the frequency error of the nuclear clock can be expressed as
\begin{equation}
\delta f =  \frac{G(\Gamma'+\delta_m^2/\Gamma')^2\sqrt{(1-e^{-x})\left(1-\alpha \right)}}
{2 R \delta_m \sqrt{5 N_{eff}}\left(
\alpha(1-e^{-x})+\left(1-\alpha\right)x e^{-x}\right)} \label{eq:41}
\end{equation}
where $x=G\theta$ and $\alpha=\gamma/G$.
To find the optimal value for $\theta$ we minimize (\ref{eq:41}) with respect to $x$ which leads to the transcendental equation for $x$:
\begin{equation}
e^x-1 =  \frac{(1-\alpha)x}
{\alpha+2 (1-\alpha) (1-x)} \,. \label{eq:42}
\end{equation}
This equation yields the optimal value of $\theta=x/G$ for the case of long total interrogation time. We again consider the cases of weak and strong interrogation field.

If the laser field is weak, i.e., if $R\ll\gamma$, $\alpha$ is close to 1. Then $1-\alpha=5R/(2\gamma (1+\delta_m^2/\Gamma'^2))$ and (\ref{eq:42}) transforms to
\begin{equation*}
e^x-1 = \frac{x}
{\frac{2 \gamma}{5 R} \left(1+\frac{\delta_m^2}{\Gamma'^2}\right)+1 - 2 x} \,.
\end{equation*}
This equation contains a large term proportional to $\gamma/R$ on the right side of the denominator which has to be compensated by another large term  $2 x_0 - 1$ to yield a solution $x_0$ of (\ref{eq:42}).
We obtain
\begin{equation}
x_0 \simeq \frac{\gamma}{5R}\left(1+\frac{\delta_m^2}{\Gamma'^2}\right)+1 \,. \label{eq:43}
\end{equation}
Therefore, for the case of weak laser field $G \theta \sim x_0 \gg 1$ and we can neglect $e^{-G \theta}$ in (\ref{eq:41}). Optimization for $\delta_m$ yields
\begin{equation}
\delta_m=\frac{\Gamma'}{\sqrt{2}},  \quad \mathrm{and}   \label{eq:44}
\end{equation}
\begin{equation}
\delta f =\frac{1}{2}\left(\frac{3}{2}\right)^{3/2}\Gamma' \sqrt{\frac{\gamma}{R\, N_{eff}}}\, . \label{eq:45}
\end{equation}

We now consider the case where the laser field is strong, i.e. when $R \gg \gamma$. We then can set $\alpha=0$ in (\ref{eq:42}). We obtain:
\begin{equation}
\theta = \frac{x_0}{G} = \frac{0.643798}{G} \,. \label{eq:46}
\end{equation}
In this case the optimal working point $\delta_m$ and the frequency fluctuation $\delta f$ are:
\begin{equation}
\delta_m=\Gamma',  \quad \mathrm{and}   \label{eq:47}
\end{equation}
\begin{equation}
\delta f =\frac{2.2778 \, \Gamma'}{\sqrt{N_{eff}}}\, . \label{eq:48}
\end{equation}
The last expression determines the ultimate stability limit that can in principle be attained by the fluorescence spectroscopy method in a single interrogation cycle for a given quantum discriminator. It is valid for a large range of parameters as discussed in Appendix~B. The optimized value for $\delta f$ (\ref{eq:48}) is exclusively determined by the decoherence rate $\Gamma'$ (which is estimated in Appendix~A) and the number of interrogated nuclei, it  does not depend on the exact values of the isomer transition energy or lifetime. To reach this performance level, the conditions $R\ll \Gamma'$ and $R\ll \Gamma_{gg'}$ have to be fulfilled, otherwise an accumulation of populations in quantum states not interacting with the laser field will decrease of $N_{eff}$ over time. However, the very large ratio of $\Gamma'/\gamma \sim 10^{6}$ gives a broad range of parameters where $R\gg \gamma$ and $R \ll \Gamma', \Gamma_{gg'}$ can be satisfied simultaneously. 

Note that in the limit of long interrogation time, the equilibrium value of the excited state population is attained in a short fraction of the interrogation cycle and the optimal excitation time takes only a small fraction of the interrogation cycle. In contrast, in the limit of short interrogation time, the equilibrium value can not be attained in one interrogation cycle and the optimal excitation time $\theta$ should be equal to the photon collection time $\theta'$.


\section{Performance of a nuclear clock based on Thorium-doped Calcium fluoride}
\label{sec:numeric_performance}
In this section we present numerically optimized relative irradiation times $t_R$, ideal working points $\delta_m$, and hence ultimate frequency performances of the solid-state nuclear clock scheme for different excitation rates $R$ and interrogation times $t$. We use the specific system parameters for $^{229}$Thorium ions doped into CaF$_2$ crystals introduced in the previous sections: $\gamma=\gamma_{s}n^3=7.8\cdot 10^{-4}$ s$^{-1}$, $\Gamma'=150\,\mathrm{Hz}\cdot2\pi=942.5$ s$^{-1}$, $\omega=2\pi \cdot 1900\, \mathrm{THz}=1.19\cdot 10^{16}$ s$^{-1}$. We suppose that the total number of Thorium nuclei is $N_{Th}=10^{14}$, the solid angle is $\Omega=4\pi/10$ and the quantum efficiency of photodetector is $k=0.1$ (which is typical for photomultipliers in our wavelength range). We obtain $N_{eff}=10^{12}$, see (\ref{e:19}). 
\begin{figure}[b]
\begin{center}
\resizebox{0.5\textwidth}{!}{ \includegraphics{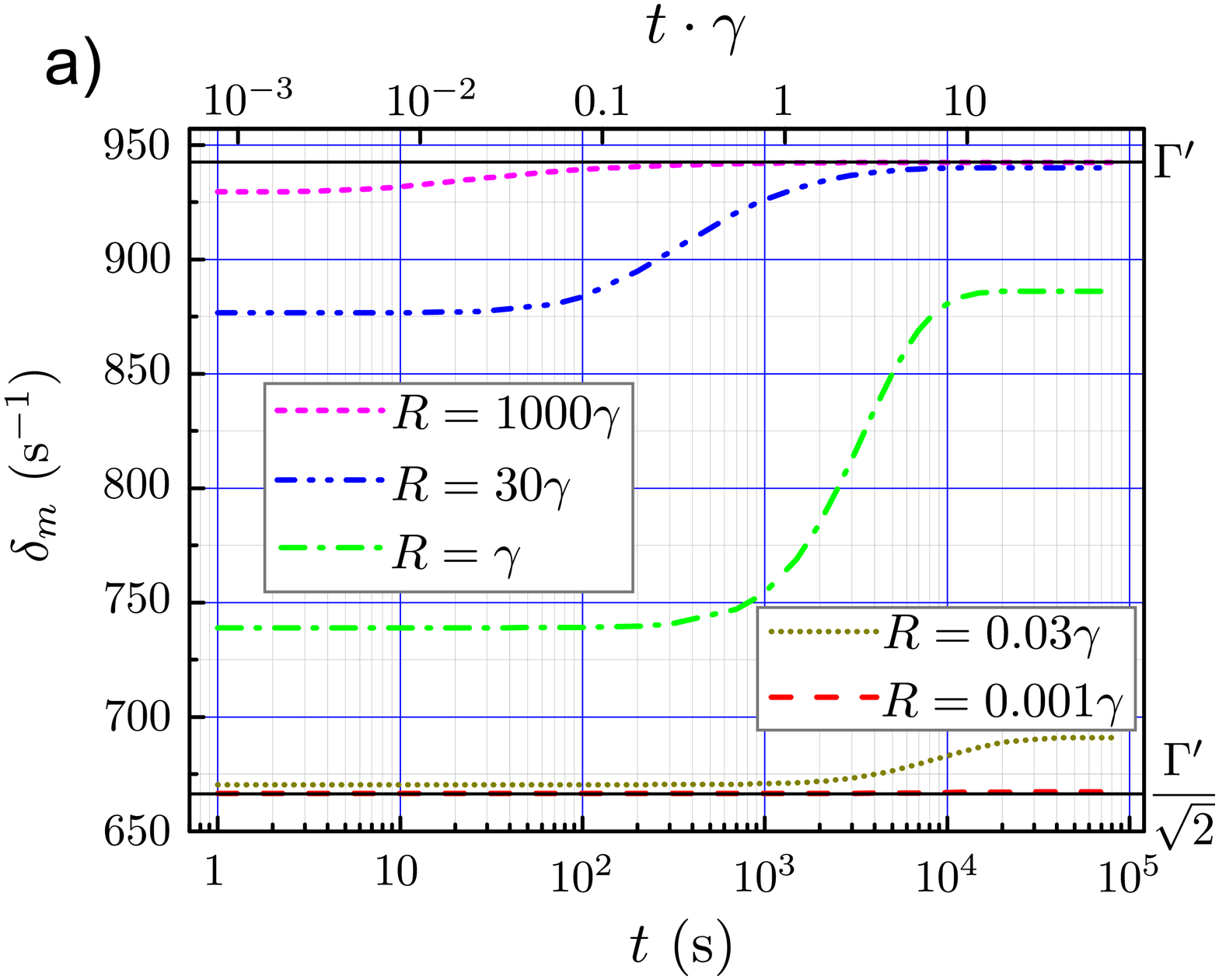}}
\hfill 
\resizebox{0.47\textwidth}{!} { \includegraphics{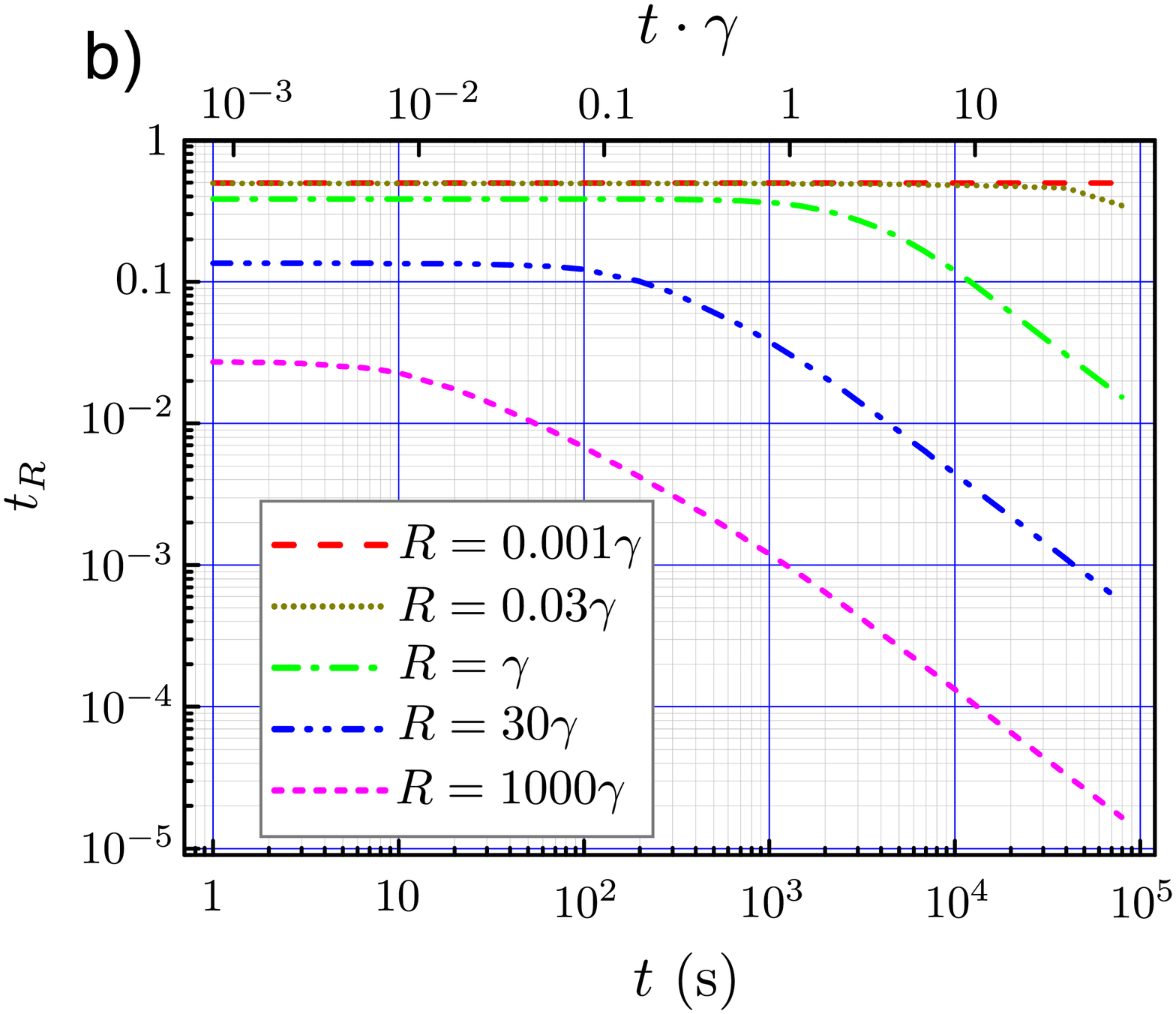}}
\end{center}
\def\baselinestretch{1.1}
\caption{Dependence of the optimal working point $\delta_m$ (a) and of $t_R$ (b) on the total interrogation time $t$ for different values of the excitation rate $R$.}
\def\baselinestretch{1.5}
\label{fig:f5}
\end{figure}

The optimized parameters $\delta_m$ and $t_R$ are represented in Figure~\ref{fig:f5} for different excitation rates $R$. One can see that for all interrogation times, the optimal working points $\delta_m$ lie between $\Gamma'$ and $\Gamma'/\sqrt{2}$ which are the limiting results for $R\ll \gamma$ (\ref{eq:37},  \ref{eq:44})  and $R \gg \gamma$ (\ref{eq:39},  \ref{eq:47}) repectively.

In general, the ideal working point is close to the inverse decoherence rate of the order kilohertz (see Appendix~A). This means that the linewidth of the interrogation laser has to be significantly narrower than this, otherwise feedback stabilization to the nuclear transition will fail. Also one can see that the optimal value of $t_R$ decreases with decreasing $t$ and increasing $R$ which means a shorter excitation than detection time. This is connected with the fact that the optimal illumination time $\theta$ should not significantly exceed $1/G(R,\delta_m)$, i.e. the typical timescale to approach the equilibrium population of the excited state $\rho_{exc}^{eq}=\frac{2}{5}(1-\gamma/G(R,\delta_m))$ according to (\ref{e:15}).

\begin{figure}
\begin{center}
\resizebox{0.5\textwidth}{!}{ \includegraphics{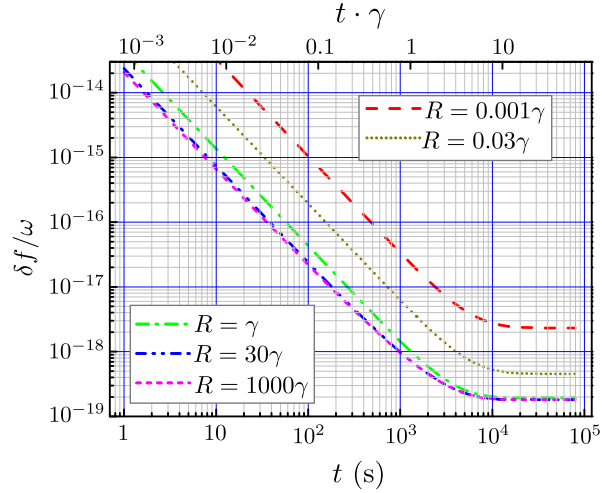}}
\end{center}
\def\baselinestretch{1.1}
\caption{Fractional frequency error $\delta f/\omega$ of the nuclear clock for different excitation rates $R$ and different total interrogation times $t$.}
\def\baselinestretch{1.5}
\label{fig:f6}
\end{figure}

Let us now turn to the nuclear clock error per one interrogation cycle obtained for the optimized parameters $\delta_m$ and $t_R$, expressed as fractional frequency error in Figure~\ref{fig:f6}. For short interrogation times, i.e., $t<0.1 \, \gamma^{-1}$ the fractional frequency error scales as $\delta f/\omega \propto t^{-3/2}$ in accordance with (\ref{eq:38}) and (\ref{eq:40}). For long interrogation times, i.e.  $t>10 \, \gamma^{-1}$, the frequency error $\delta f/\omega$ becomes independent of the interrogation time in accordance with (\ref{eq:45}) and (\ref{eq:48}). We note also that for strong pumping, i.e. $R>30\,\gamma$, the clock performance is practically independent of the exact value of $R$, for long interrogation times this is already the case for $R>\gamma$. For $t>10/\gamma$ and $R$ more than a few $\gamma$, the fractional frequency error $\delta f/\omega$ per one interrogation cycle is essentialy independent of $R$ and $t$ and is equal to the ultimate value of about $1.8\cdot 10^{-19}$ in correspondance with (\ref{eq:48}).

\begin{figure}
\begin{center}
\resizebox{0.48\textwidth}{!}{ \includegraphics{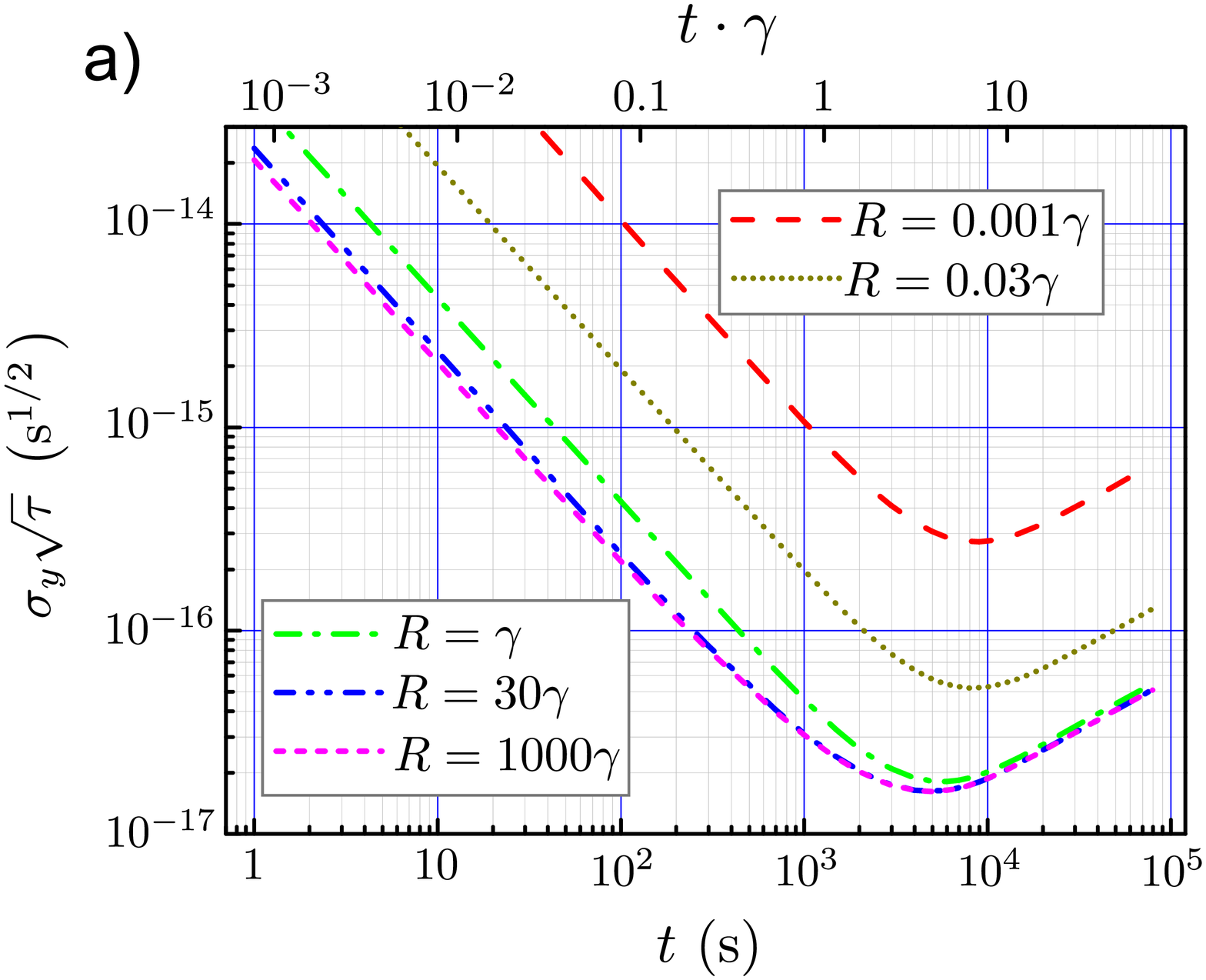}}
\hfill 
\resizebox{0.48\textwidth}{!} { \includegraphics{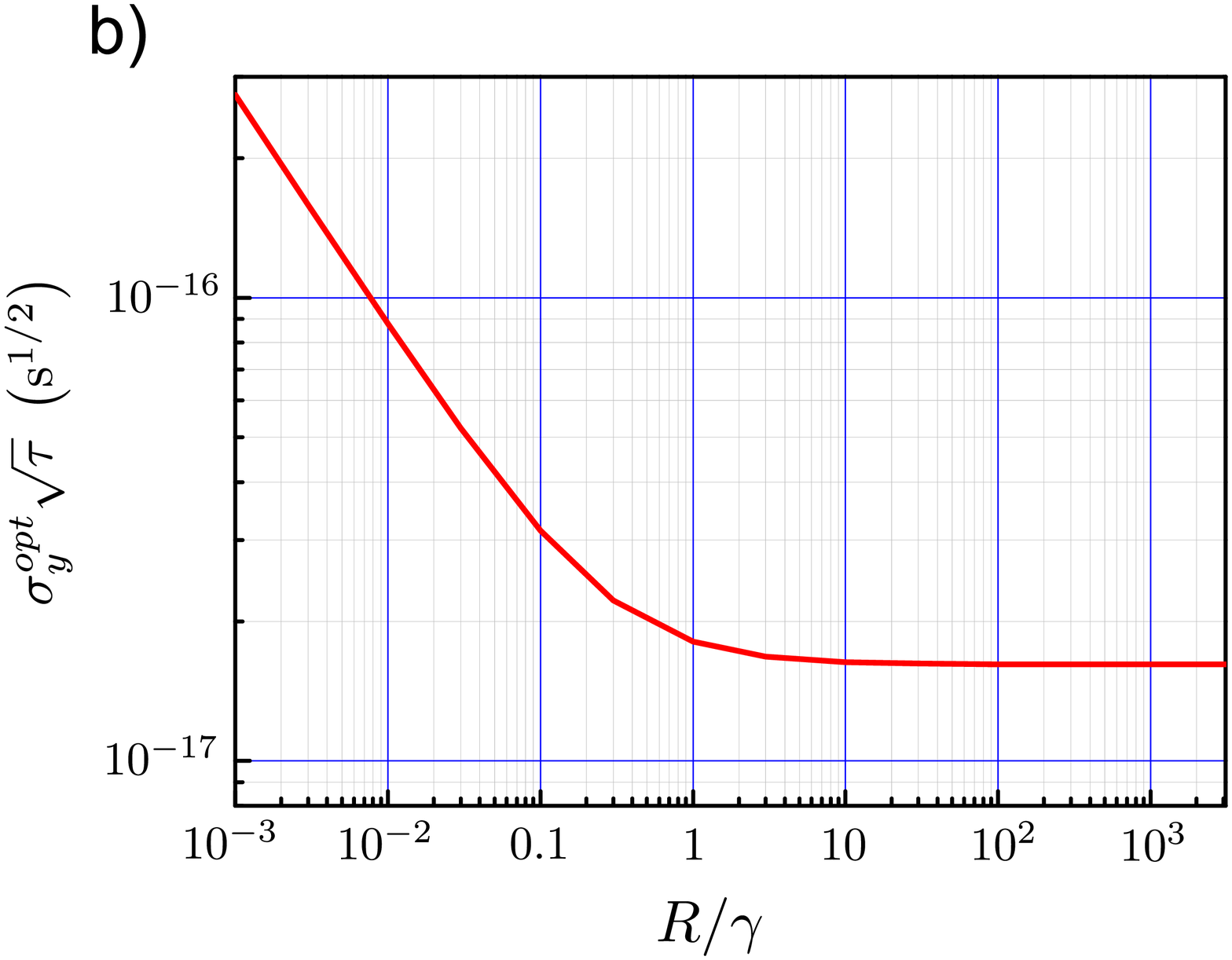}}
\end{center}
\def\baselinestretch{1.1}
\caption{Nuclear clock fractional instability $\sigma_y \sqrt{\tau}=\delta f/\omega \cdot \sqrt{t}$ as a funtion of total interrogation time $t$ for different excitation rates $R$ (a) and the ultimate clock instability $\sigma_y^{opt} \sqrt{\tau}$ obtained for optimal value of $t$, as a function of $R$ (b).}
\def\baselinestretch{1.5}
\label{fig:f7n}
\end{figure}

We now consider the parameter $\delta f/\omega \cdot \sqrt{t}$ which describes the \emph{fractional clock instability}. It should be optimized with respect to interrogation time $t$ if one wants to perform a precise measurement of the frequency of the isomer transition (in the solid-state environment) by a long-term series of interrogations. It is also the relevant quantity describing the instability of the solid-state nuclear frequency standard when a narrow band laser is feedback-stabilized to the transition and compared to a primary frequency standard. If the series of measurements takes a time $\tau=N\cdot t$, then the fractional instability $\sigma_y(\tau)$ of the series is equal to the fractional frequency error averaged over the series. In shot noise limit we obtain
\begin{equation}
\sigma_y(\tau)= \frac{\delta f}{\omega\sqrt{N}}=\frac{\delta f\sqrt{t}}{\omega\sqrt{\tau}}. \label{eq:49s}
\end{equation}
Hence, to reduce $\sigma_y(\tau)$ one should optimize (minimize) $\delta f\sqrt{t}/\omega=\sigma_y\sqrt{\tau}$. We find that the optimal value for $t$ is of the order of a few $\gamma^{-1}$, only weakly depending on the excitation rate $R$, see Figure~\ref{fig:f7n}\,(a). The ultimate clock instability is given by the minimum value of $\sigma_y\sqrt{\tau}$ at the optimal value of $t$. It saturates at $\approx 1.63\times 10^{-17}$ s$^{1/2}$ for high excitation rates ($R/\gamma\gtrsim 10$) and scales as $R^{-1/2}$ for lower excitation rates ($R/\gamma\lesssim 0.1$), see Figure~\ref{fig:f7n}\,(b). The resonance intensity $I_0$ of the laser source required to attain the ``high excitation rate regime'' $R/\gamma\gtrsim 10$ can be estimated using (\ref{e:16}). For the system parameters used in this section it yields $I_0\simeq 16\, \mathrm{mW/cm^2}$. Other optimized parameters corresponding to $R=10\gamma$ are: $t\simeq 3.88/\gamma \simeq 5\cdot 10^3$ s, $t_R \simeq 0.026$ and $\delta_m\simeq 0.99 \Gamma'$.

A few words on the validity of approximation~(\ref{uneq:9}) and the optimal value of the temperature: taking $R=10 \gamma$ and using (\ref{e:16}), we easily estimate the ``reasonable" value of $V_{eg}$ as $V_{eg} \simeq 2\pi \cdot 0.53\, \mathrm{Hz}$.
The mixing rate $\Gamma_{gg'}$ should be large in comparison with this value. Our estimation (see section~\ref{sec:effects}) gives $\Gamma_{gg'} \simeq 2 \pi \cdot (5 - 100)$\,Hz for room temperature, and $\Gamma_{gg'} \propto T^2$, then we can take $\Gamma_{gg'} [\mathrm{Hz}] \simeq (5\cdot 10^{-5} - 10^{-3})\cdot T^2\,[K]$. Condition~(\ref{uneq:9}) gives $T[K] > \sqrt{\frac{0.53}{5\cdot 10^{-5}-10^{-3}}}, \quad \mathrm{or} \quad T > (20 - 100) \,\mathrm{K}$. If (\ref{uneq:9}) is not fulfilled, in a real multilevel system a significant fraction of nuclei will be pumped into the dark state, and the growing of the excitation rate $R$ will reduce the performance. On the other hand, even at liquid nitrogen temperatures the contribution of the temperature-dependent 2nd order Doppler broadening term to the total decoherence rate $\Gamma'$ becomes comparable to the temperature-independent magnetic dipole interaction term, and increasing the temperature is undesirable. We expect that the optimal sample temperature should be around the liquid nitrogen temperature. A more detailed analysis requires more precises data on the quadrupole relaxation and is beyond the scope of this article. Nevertheless we can claim that the optimal excitation rate $R$ should be of the order of a few $\gamma$.

In an experimental implementation, the performance of the solid-state nuclear clock will be limited by technical constrains such as the temperature stability of the sample, the frequency and intensity stability of the interrogation laser system, and noise of the detection system. If these limitations can be overcome, the solid-state approach can reach the $10^{-19}$ fractional instability level within a few hours of averaging. It may hence be competitive with the trapped-ion scheme~\cite{Campbell11} concerning stability for a (temperature stabilized) liquid nitrogen temperature sample. Note that the fluorescence stabilization scheme developped here is opperated continuously without deadtime between cycles as no state initialization or re-trapping of particles is required. The global line shifts (several hundreds of MHz) due to crystal fields will obviously degrade the clock accuracy and require calibration with a primary standard.

\section{Conclusion}
\label{sec:conclusion}
We have analyzed the possibility of constructing a ``solid-state nuclear clock'' from $^{229}$Thorium nuclei implanted into VUV transparent Calcium fluoride crystals. We expect static global line shifts to be dominated by the temperature-dependent electric monopole and electric quadrupole interactions and by 2$^{nd}$ order Doppler shifts, necessitating active temperature stabilization of the sample. The electric quadrupole interaction leads to a splitting of the nuclear ground and isomer levels of the order of a few hundred MHz and is the main source of state mixing. As broadening effects we considered again the 2$^{nd}$ order Doppler shift, the electric quadrupole, and the magnetic dipole interaction. The magnetic dipole interaction does not depend on temperature and is hence expected to be the dominating source of decoherence (followed by the 2$^{nd}$ order Doppler effect) at liquid Nitrogen temperatures. For the CaF$_2$ crystal we estimate a decoherence rate of 150\,Hz. The extreme gap between this rate and the nuclear state lifetime renders the application of commonly used coherent interrogation schemes (i.e. Rabi, Ramsey) inappropriate. We therefore developed an alternative clock stabilization scheme based on fluorescence spectroscopy. A detailed analysis reveals a favorable scaling with the number of interogated nuclei, validating the solid-state approach.
 Neglecting all possible technical constrains (i.e. frequency and intensity stability of the interrogation laser), supposing that the sample contains $10^{14}$ Thorium nuclei and that the detection probability is 1\%, we find an ultimate fractional instability limit of $1.8\cdot 10^{-19}$ for a single interrogation cycle taking about $10^{4}$\,s.
\ack

This research was supported by the Austrian Science Fund (FWF) by projects M1272-N16 (TheoNAC) and Y481-N16; by the European Research Council (ERC): starting grant 258603 NAC, by the State Foundation of Fundamental Researches of Ukraine (project F40.2/039), by the Russian Federal Program ``Scientific and scientific-pedagogical personnel of innovative Russia in 2009-2013'', by the Russian
Foundation for Basic Research in grant RFBR--11--02--90426\_Ukr\_f\_a, and by the grant of President of Russian Federation for young candidates of science, project MK-5318.2010.2. We thank P. Mohn, P. Dessovic, and R. Jackson for helpful discussion and calculations concerning the microscopic structure of the Thorium doping complex.

\section*{Appendix A. Decoherence and mixing due to magnetic dipole interaction}
\label{sec:relaxations}

A Calcium fluoride crystal is formed from $^{19}$F Fluorine and a natural mixture of Calcium isotopes ($^{40}$Ca, $^{42}$Ca, $^{43}$Ca, $^{44}$Ca, and $^{46}$Ca). The only Ca isotope with non-zero nuclear magnetic moment $^{43}$Ca has a small abundance of the order of 0.14\,\% whereas the nucleus of the only abundant $^{19}$F isotope has a finite magnetic moment $\mu_{F}=2.63\, \mu_N$ and a nuclear spin $s=1/2$. In the following we consider only the magnetic moment of the $^{19}$F nuclei in the CaF$_2$ lattice to interact with the Thorium nuclear transition. The relaxation of the nuclear magnetic moment of Thorium (or Fluorine) nuclei is caused by the interaction with a random magnetic field $\mathcal{\vec{H}}$ at the position of the considered nucleus, created by the surrounding Fluorine nuclei.

For the calculation of the relaxation and mixing rates we use the formalism of relaxation theory presented in~\cite{Vanier}. Here we briefly describe this formalism and apply it first to the magnetic moment of Fluorine nuclei. This yields the characteristic rate of Fluorine spin-spin relaxation and therefore the characteristic correlation time of $\mathcal{\vec{H}}$. Then we calculate the relaxation rate for the ultraviolet transitions between different quadrupole sublevels of the ground and isomer state of the Thorium nucleus.

\subsection*{Relaxation method}
Here we follow a method developed in~\cite{Vanier} for the calculation of the effect of a stationary random perturbation on the elements of the density matrix. The evolution of the density matrix is described by the Liouville-von Neumann equation
\begin{equation}
\frac{d\hat{\rho}}{dt}=-\frac{i}{\hbar}\left[\hat{H},\hat{\rho}\right], \quad \mathrm{where} \quad{\hat{H}=\hat{H}_0+\hat{H}_1^C+\hat{H}_1}. \label{eq:49}
\end{equation}
Here $[\hat{A},\hat{B}]\equiv \hat{A}\hat{B}-\hat{B}\hat{A}$ is a commutator, $\hat{H}_0=\hbar \sum_\alpha |\alpha\rangle\,\omega_{\alpha}\langle \alpha |$ is the Hamiltonian of the unperturbed system, $\hat{H}_1^C$ is the Hamiltonian of interaction with the coherent external field and $\hat{H}_1$ is the Hamiltonian of interaction with a random field responsible for the relaxation, in our case with the random magnetic field $\mathcal{\vec{H}}$. In order to eliminate the strong contribution $\hat{H}_0$ from (\ref{eq:49}) one applies the unitary transformation $\hat{L} \rightarrow \hat{L}'= e^{i\hat{H}_0t/\hbar}\hat{L}e^{-i\hat{H}_0t/\hbar}$ to all the operators in (\ref{eq:49}) (interaction representation). We obtain the equation for the density matrix:
\begin{equation}
\frac{d\hat{\rho}'}{dt}=-\frac{i}{\hbar}\left[\hat{H}_1^{C\prime},\hat{\rho}'\right]-
\frac{i}{\hbar}\left[\hat{H}_1',\hat{\rho}'\right].  \label{eq:50}
\end{equation}
The solution of this equation can be written as
\begin{equation}
\hat{\rho}'(t)=\hat{\rho}'(0)-\frac{i}{\hbar}\int_0^{t}\left[(\hat{H}_1^{C\prime}+\hat{H}_1'),\hat{\rho}'(t')\right]dt'.  \label{eq:51}
\end{equation}
Substituting (\ref{eq:51}) into the second term on the right-hand side of (\ref{eq:50}) yields:
\begin{eqnarray}
\frac{d\hat{\rho}'}{dt}&=&-\frac{i}{\hbar}\left[\hat{H}_1^{C\prime}(t),\hat{\rho}'(t)\right]-
\frac{i}{\hbar}\left[\hat{H}_1^{\prime}(t),\hat{\rho}'(0)\right] \nonumber \\
&&-\frac{1}{\hbar^2}\int_0^t \left[ \hat{H}_1^{\prime}(t),
\left[ \bigg(\hat{H}_1^{C\prime}(t')+\hat{H}_1^{\prime}(t')\bigg),\hat{\rho}'(t')\right] \right] dt'.  \label{eq:52}
\end{eqnarray}
We assume that the characteristic time scale of the correlations of fluctuation of the random field $\hat{H}_1^{\prime}(t)$ is much smaller than all other characteristic time scales, including the characteristic time scale of variation of the density matrix \footnote{Although this condition is not entirely fulfilled for the relaxation of Fluorine spins in the random magnetic field $\vec{\mathcal{H}}$ created by neighboring spins, quite good correspondance with experimental results indicates the applicability of this theory. For the Thorium nucleus interacting with $\vec{\mathcal{H}}$, this condition is fulfilled much better due to the much lower magnetic moment compared to Fluorine.}.
Then one can choose the origin in such a way that the density matrix would not change noticeable during this time interval (one may substitute $\rho'(t')$ by $\rho'(t)$) but all the correlations of the random field will disappear (one may take $-\infty$ as a lower limit of integration). Also suppose that all the matrix elements of $\hat{H}_1^{\prime}(t)$ have zero mean values (otherwise we would add them to $\hat{H}_1^{C\prime}$). Performing the averaging over the ensemble (indicated by overline), we obtain
\begin{equation}
\frac{d\hat{\rho}'}{dt}=-\frac{i}{\hbar}\left[\hat{H}_1^{C\prime}(t),\hat{\rho}'(t)\right]
-\frac{1}{\hbar^2}\int_{-\infty}^t \left[\, \overline{\hat{H}_1^{\prime}(t),
\bigg[ \hat{H}_1^{\prime}(t')},\hat{\rho}'(t)\bigg] \right] dt'.  \label{eq:53}
\end{equation}
The first term in (\ref{eq:53}) corresponds to the evolution of the density matrix under the action of a coherent field. The second term describes the relaxation. If the perturbations are stationary, it can be written as
\begin{equation}
\mathbb{R}\hat{\rho}'(t)\equiv -\frac{1}{\hbar^2}\int_0^{\infty} \left[\, \overline{\hat{H}_1^{\prime}(t),
\bigg[ \hat{H}_1^{\prime}(t-\theta)},\hat{\rho}'(t)\bigg] \right] d\theta.  \label{eq:54}
\end{equation}
Expanding the commutators in (\ref{eq:54}), one obtains:
\begin{equation}
\left(\mathbb{R}\hat{\rho}'\right)_{\alpha \alpha'}=\sum_{\beta \beta'} R_{\alpha \alpha' \beta \beta'} \, \rho'_{\beta \beta'}\, \exp[i(\omega_{\alpha \beta}-\omega_{\alpha' \beta'})t],  \label{eq:55}
\end{equation}
where $\omega_{ij}=\omega_{i}-\omega_{j}$,
\begin{eqnarray} 
R_{\alpha \alpha' \beta \beta'}&=&\frac{1}{\hbar^2}\Bigg( 
j_{\alpha \beta \alpha' \beta'}(\omega_{\alpha \beta})+
j_{\beta' \alpha' \beta \alpha}(\omega_{\beta' \alpha'}) \nonumber \\
& 
-&\delta_{\alpha \beta} \sum_{\gamma}
j_{ \beta' \gamma \alpha' \gamma} (\omega_{\beta' \gamma})
-\delta_{\alpha' \beta'} \sum_{\gamma}
j_{\gamma \beta \gamma \alpha} (\omega_{\gamma \beta})
 \Bigg),  \label{eq:56}
\end{eqnarray}
and
\begin{equation}
j_{\alpha \beta \alpha' \beta'}(\omega)=
\int_0^{\infty} \overline{\hat{H}_{1\alpha \beta}(t)
\hat{H}_{1\beta' \alpha'}(t-\theta)}\, e^{-i\omega\theta}\, d\theta  \label{eq:57}
\end{equation}
(note that in the right-hand side of the last expressions, the matrix elements of $\hat{H}_1$ are taken in the ``\,initial'' Schr\"odinger representation). Expression (\ref{eq:55}) for the relaxation operator $\mathbb{R}$ contains oscillating terms that can be neglected. As a result we keep only the terms where $\omega_{\alpha\beta}=\omega_{\alpha'\beta'}$. 
Except for some special cases, this condition is fulfilled only if $\alpha=\alpha'$ and $\beta=\beta'$ (mixing of populations), or if $\alpha=\beta$ and $\alpha'=\beta'$ (relaxation of coherences).

Using (\ref{eq:54}) --- (\ref{eq:57}) one can calculate the relaxation and population mixing rates.

\subsection*{Spin relaxation of a Fluorine nucleus}

Let us consider a single Fluorine nucleus. The spin of this nucleus can take one of two projections $m=\pm 1/2$ with respect to the arbitrary quantization axis $z$. Let us denote $|1\rangle=|m=-1/2\rangle$ and $|2\rangle=|m=1/2\rangle$. The magnetic moment operator $\hat{\vec{\mu}}$ for the spin of this nucleus can be expressed as $\hat{\vec{\mu}}=\hat{\vec{s}}\, \mu_F/s$. Therefore we can express the operator $\hat{H}_1=-\hat{\vec{\mu}} \cdot \vec{\mathcal{H}}$ as
\begin{eqnarray}
\hat{H}_1&=&\mu_F \bigg(\mathcal{H}_z\left(|1\rangle \langle 1|-|2\rangle \langle 2| \right) \nonumber \\
&-&\mathcal{H}_x\left(|1\rangle \langle 2|+|2\rangle \langle 1| \right)-
i \mathcal{H}_y\left(|1\rangle \langle 2|-|2\rangle \langle 1| \right)\bigg).  \label{eq:58}
\end{eqnarray}

In this work we use the ``diffusion'' model for the correllations of the random magnetic field:
\begin{equation}
\left\langle\mathcal{\vec{H}}(t) \right\rangle=0, \; 
\left\langle\mathcal{\vec{H}}^2(t) \right\rangle=\mathcal{H}_0^2, \;
\left\langle\mathcal{H}_i(t)\mathcal{H}_j(t-\theta) \right\rangle=\delta_{ij}\frac{\mathcal{H}_0^2}{3}\, e^{-|\theta|/t_c}. \label{eq:59}
\end{equation}
The average square of the random magnetic field $\mathcal{H}_0^2$ and the correlation time $t_c$ will be determined later. 

According to (\ref{eq:56}) the relaxation rate $R_{1212}$ of the coherence $\rho_{12}$ is
\begin{eqnarray}
R_{1212}&=&\frac{1}{\hbar^2} \left[ j_{1122}(0)+j_{2211}(0)-j_{2121}(0) \right. \nonumber \\
&&-\left. j_{2222}(0)-j_{1111}(0)-j_{2121}(0)  \right]. \label{eq:60}
\end{eqnarray}
From (\ref{eq:57}) and (\ref{eq:59}) one finds
\begin{eqnarray}
R_{1212}&=&\frac{1}{\hbar^2} \left[ j_{1122}(0)+j_{2211}(0) \right. \nonumber \\
&&-\left. j_{2222}(0)-j_{1111}(0)-2 j_{2121}(0)  \right]. \label{eq:61}
\end{eqnarray}
where
\begin{eqnarray}
j_{1111}(0)&=&j_{2222}(0)=-j_{1122}(0)=-j_{2211}(0)=\frac{\mu_F^2 \mathcal{H}_0^2}{3}t_c 
\nonumber \\
j_{2121}(0)&=&2 \frac{\mu_F^2 \mathcal{H}_0^2}{3}t_c \label{eq:62}
\end{eqnarray}
which yields the coherence relaxation rate
\begin{equation}
R_{1212}=-\frac{8}{3}\frac{\mu_F^2 \mathcal{H}_0^2 t_c}{\hbar^2}. \label{eq:64}
\end{equation}
Similarily one finds the population mixing rate as
\begin{eqnarray}
R_{1111}&=&-R_{1122} \nonumber \\
&=&-\frac{j_{2121}(0)+j_{1212}(0)}{\hbar^2} =-\frac{4}{3}\frac{\mu_F^2 \mathcal{H}_0^2 t_c}{\hbar^2}. \label{eq:65}
\end{eqnarray}

Now we can estimate the correlation time $t_c$ which is determined by the population relaxation rate:
\begin{equation}
t_c=\sqrt{\frac{3 \hbar^2}{4\mu_F^2 \mathcal{H}_0^2}}. \label{eq:66}
\end{equation}
To find $\mathcal{H}_0^2$ we recall that in CaF$_2$ the Fluorine nuclei form a simple qubic lattice with a distance of $a/2=2.73$ \AA \,  between two neighbors. Spins of different Fluorine nuclei are not correlated, therefore the average square $\mathcal{H}_0^2$ of the random magnetic field $\vec{\mathcal{H}}$ is the sum of squares of magnetic fields produced by different Fluorine nuclei:
\begin{equation}
\mathcal{H}_0^2=\sum_a\mathcal{H}_{a}^2. \label{eq:67}
\end{equation}
$\vec{\mathcal{H}}_a$ is the magnetic field created by $a$-th Fluorine nucleus whose radius-vector is $\vec{r}_a$ with respect to the nucleus of interest. Using $\hat{\vec{\mu}}_a=2 \hat{\vec{s}}_a\, \mu_F$, and $\mathcal{\vec{H}}_a=(3\vec{r} (\hat{\vec{\mu}}_a\vec{r}_a) - \hat{\vec{\mu}}_a r_a^2)/r_a^5$, we find
\begin{equation}
\mathcal{H}_{a}^2=\frac{3 (\hat{\vec{\mu}}_a\vec{r}_a)^2+\hat{\vec{\mu}}_a^2r_a^2}{r_a^8}
= \frac{6\mu_F^2}{r_a^6}.
\label{eq:68}
\end{equation}
Therefore
\begin{equation}
\mathcal{H}_{0}^2=\frac{6\mu_F^2}{r_a^6} \sum\limits_{\scriptstyle{l,m,n=-\infty}}^{\infty} \left[l^2+m^2+n^2\right]^{-3}\simeq \frac{6\mu_F^2}{r_a^6} \cdot 8.4
\label{eq:69}
\end{equation}
(here the sum is taken over $l,m,n$ excluding $l=m=n=0$), or $\mathcal{H}_{0}^2\simeq 21.5$ G$^2$ which leads to a relaxation time $t_c \simeq 15 \, \mu\mathrm{s}$ according to (\ref{eq:66}). Note that this value is quite close to experimentally observed spin-spin relaxation times $\sim 20\,\mu$s~\cite{Jeener67}.

\subsection*{Spin relaxation of a Thorium nucleus}

\begin{figure}
\begin{center}
\resizebox{0.35\textwidth}{!}{ \includegraphics{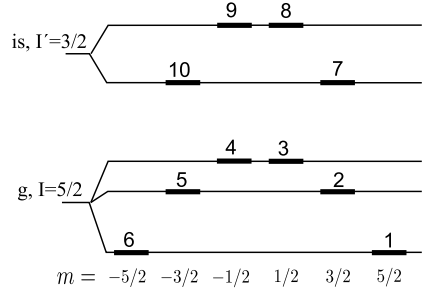}}
\end{center}
\def\baselinestretch{1.1}
\caption{Notation for Thorium nucleus spin states used in this paragraph.}
\def\baselinestretch{1.5}
\label{fig:f8}
\end{figure}

We now consider an individual Thorium nucleus interacting with random magnetic fields created by the surrounding Fluorine nuclei. The quantization axis is fixed by the direction of the main axis of the electric field gradient. Let us denote the different states of the Thorium nucleus as $|i\rangle$, where $i$ is a number from 1 to 10 corresponding the spin state shown in Figure~\ref{fig:f8}.
The average square $\mathcal{H}_{Th}^2$ of the magnetic field at the position of the Thorium nucleus is equal to
\begin{equation}
\mathcal{H}_{Th}^2=\frac{6\mu_F^2}{r_a^6} \sum \limits_{l,m,n=-\infty}^{\infty}\Bigg[\sum\limits_{k=l,m,n}\left(k-\frac{1}{2}\right)^2\Bigg]^{-3}\simeq \frac{6\mu_F^2}{r_a^6} \cdot 19\simeq 48.5 \, \mathrm{G}^2.
\label{eq:70}
\end{equation}

The relaxation rates $\Gamma_{ij}=-R_{ijij}$ for optical coherences $\rho_{ij}$ due to interaction with the random magnetic field created by surrounding Fluorine spins can be calculated in the same manner as for a single Fluorine spin above. These rates (for allowed magnetic dipole transitions) are equal to:
\begin{eqnarray}
\Gamma_{4,8}&=&\Gamma_{3,9}=\frac{\mathcal{H}_{Th}^2 t_c}{3 \hbar^2} 
\left[\mu_e^2+\frac{2}{15}\mu_e \mu_g+\frac{19}{25}\mu_g^2 \right]
\simeq 2\pi\cdot 142 \,\mathrm{Hz}, \nonumber \\
\Gamma_{3,8}&=&\Gamma_{4,9}=\frac{\mathcal{H}_{Th}^2 t_c}{3 \hbar^2} 
\left[\mu_e^2-\frac{2}{15}\mu_e \mu_g+\frac{19}{25}\mu_g^2 \right]
\simeq 2\pi\cdot 150 \,\mathrm{Hz}, \nonumber \\
\Gamma_{2,8}&=&\Gamma_{5,9}=\frac{\mathcal{H}_{Th}^2 t_c}{3 \hbar^2} 
\left[\mu_e^2-\frac{2}{5}\mu_e \mu_g+\frac{9}{25}\mu_g^2 \right]
\simeq 2\pi\cdot 84 \,\mathrm{Hz}, \nonumber \\
\Gamma_{3,7}&=&\Gamma_{4,10}=\frac{\mathcal{H}_{Th}^2 t_c}{3 \hbar^2} 
\left[\mu_e^2-\frac{2}{5}\mu_e \mu_g+\frac{19}{25}\mu_g^2 \right]
\simeq 2\pi\cdot 158 \,\mathrm{Hz}, \nonumber \\
\Gamma_{2,7}&=&\Gamma_{5,10}=\frac{\mathcal{H}_{Th}^2 t_c}{3 \hbar^2} 
\left[\mu_e^2-\frac{6}{5}\mu_e \mu_g+\frac{9}{25}\mu_g^2 \right]
\simeq 2\pi\cdot 108 \,\mathrm{Hz}, \nonumber \\
\Gamma_{1,7}&=&\Gamma_{6,10}=\frac{\mathcal{H}_{Th}^2 t_c}{3 \hbar^2} 
\left[\mu_e^2-2\mu_e \mu_g+\mu_g^2 \right]
\simeq 2\pi\cdot 251 \,\mathrm{Hz}, \label{eq:71}
\end{eqnarray}
Here we used the values of the Thorium nuclear magnetic moment in the ground state $\mu_g=0.45 \mu_N$ and in the isomer state $\mu_e=-0.076 \mu_N$ respectively~\cite{Dykhne98}. The decoherence rates $\Gamma_{3,8}=\Gamma_{4,9}$ have been used in the evaluations of the nuclear clock performance in section~\ref{sec:numeric_performance}.


\section*{Appendix B. Validity of analytic expressions for the frequency fluctuation}
\label{sec:analytic_validity}

\begin{figure}[t]
\begin{center}
\resizebox{0.48\textwidth}{!}{ \includegraphics{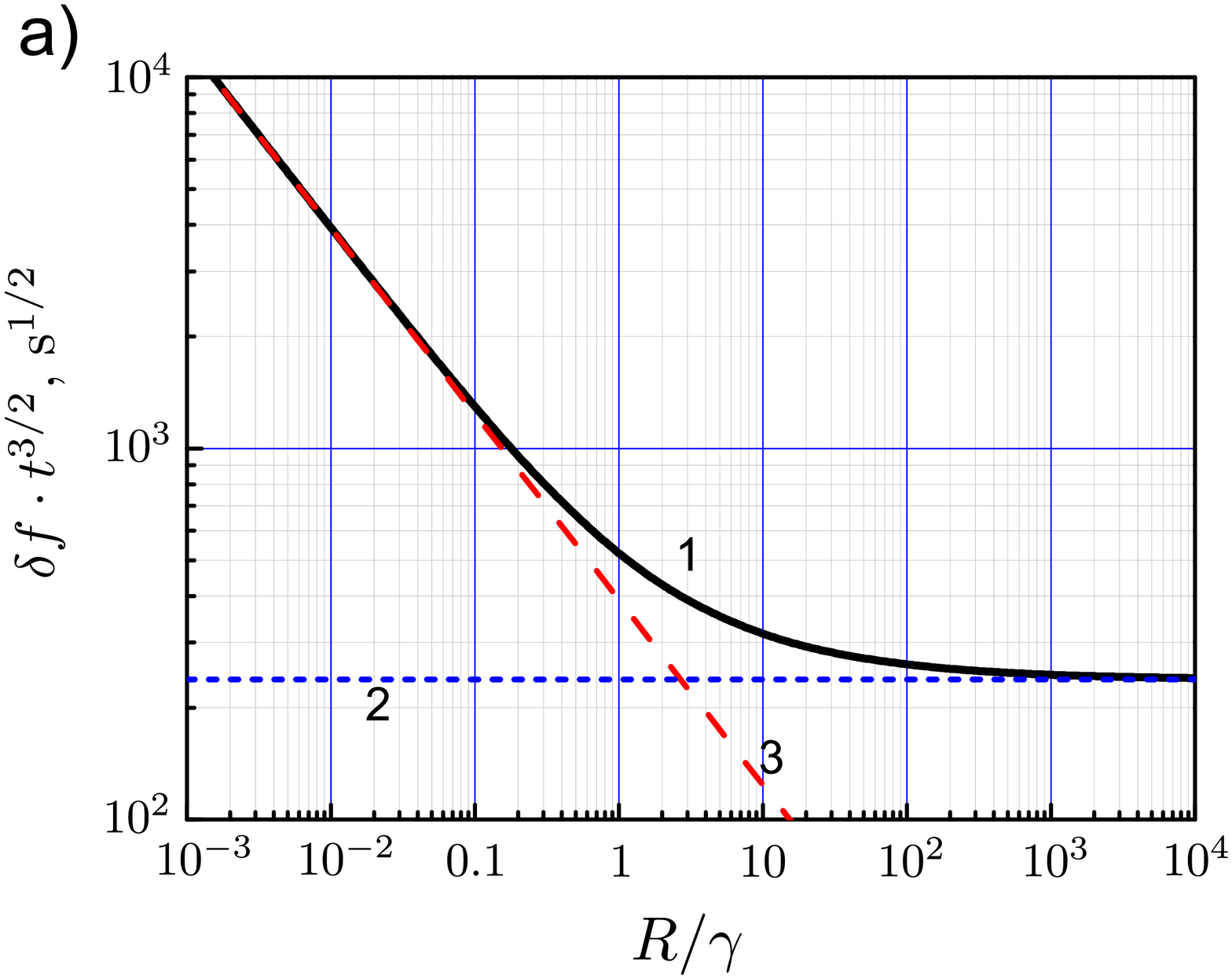}}
\hfill 
\resizebox{0.48\textwidth}{!} { \includegraphics{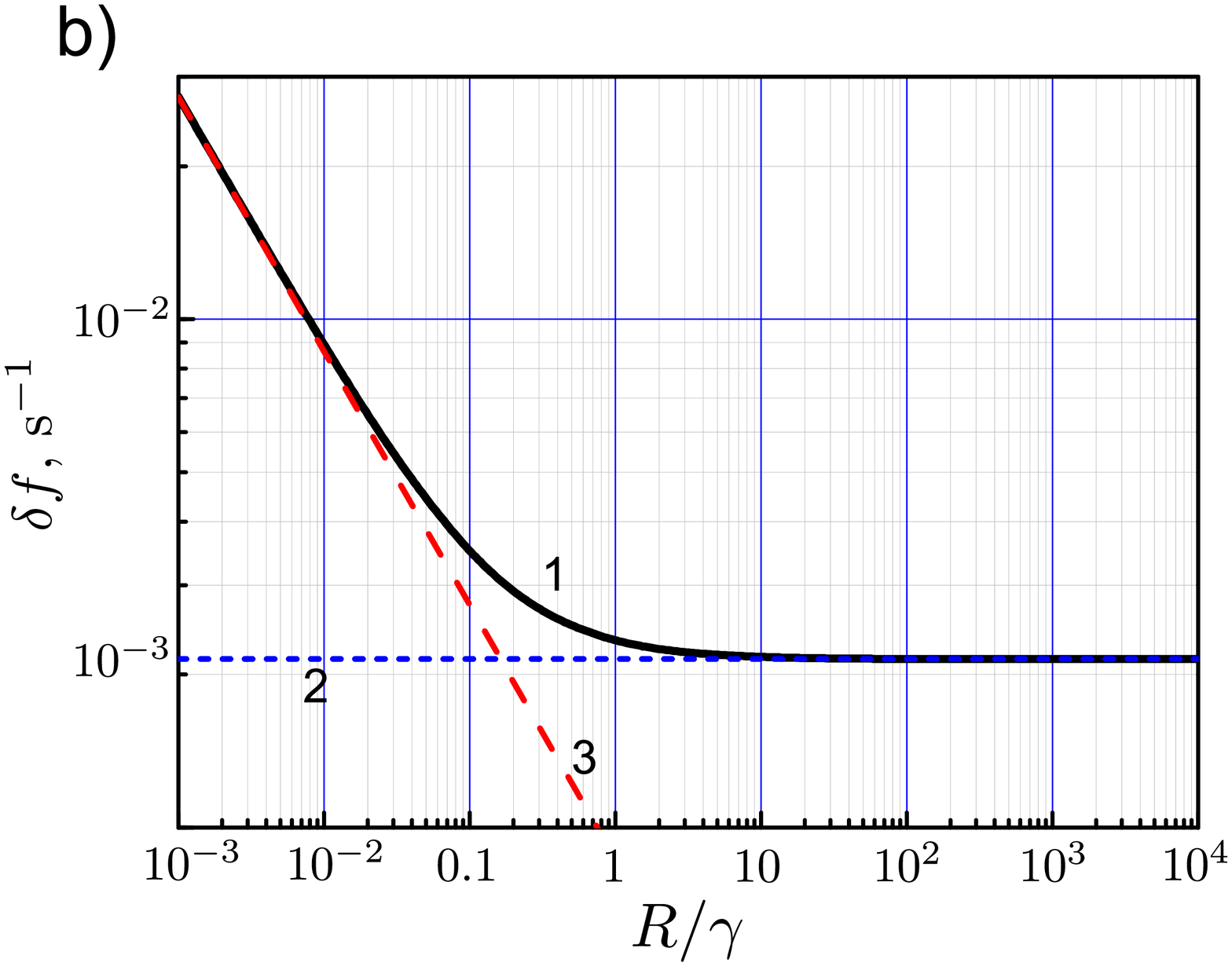}}
\end{center}
\def\baselinestretch{1.1}
\caption{(a): dependence of $\delta f \cdot t^{3/2}$ for short interrogation time $t$ on the excitation rate $R$ obtained by the optimization of (\ref{eq:32}) (curve 1), using approximate formula (\ref{eq:40}) (line 2), and (\ref{eq:38}) (line 3). (b): dependence of $\delta f$ for long $t$ on $R$ obtained by the optimization of (\ref{eq:32}) (curve 1), using approximate formula (\ref{eq:48}) (line 2), and (\ref{eq:45}) (line 3).}
\def\baselinestretch{1.5}
\label{fig:f7}
\end{figure}

Here we compare the numerically optimized results for $\delta f$ obtained in section~\ref{sec:numeric_performance} to the analytic expressions obtained in section~\ref{sec:interrogation} for the limiting cases of short and long interrogation times. For short $t$, $\delta f \propto t^{-3/2}$, therefore we express 
$\delta f \cdot t^{3/2}$ as a function of $R$, and compare theses results with the results of the analytical formulas (\ref{eq:38}) and (\ref{eq:40}), see Figure~\ref{fig:f7}(a). For long $t$ we express $\delta f$ as a function of $R$ and compare this results with formulas (\ref{eq:45}) and (\ref{eq:48}).

We conclude that the approximate analytic formulas (\ref{eq:38}), (\ref{eq:40}), (\ref{eq:45}), and (\ref{eq:48}) well describe the behavior of the optimized frequency error $\delta f$ of the nuclear clock for different limiting cases of short and long interrogation time and for weak and strong laser fields. Taking into account the quite smooth dependence of $\delta f$ on the interrogation time (see Figure~\ref{fig:f6}), we conclude that one can use expressions (\ref{eq:38}), (\ref{eq:40}), (\ref{eq:45}), and (\ref{eq:48}) as a lower-bound estimation for the frequency fluctuation $\delta f$ of the fluorescence spectroscopy interrogation scheme.


\end{document}